\documentclass{article}

\usepackage{amsfonts,amssymb,amsmath,epsfig}
\usepackage[american]{babel}
\usepackage{graphicx}
\usepackage{subcaption}

\newtheorem{theorem}{Theorem}[section]
\newtheorem{Remark}[theorem]{Remark}

\newcommand{\cB}{\mathcal{B}}

\newcommand{\RR}{\mathbb{R}}

\newcommand{\E}{\mathbb{E}}

\newcommand{\f}{\mathbf{f}}

\newcommand{\sm}{\smallskip}

\newcommand{\no}{\noindent}

\newcommand{\ve}{\varepsilon}
\newcommand{\vf}{\varphi}
\newcommand{\si}{\sigma}

\font\dc=cmbxti10 

\begin{document}

\title{Stochastic Evolving Differential Games Toward a Systems Theory of Behavioral Social Dynamics}

\author{
Giulia Ajmone Marsan \\
\small Organization for Economic Co-Operation and Development\\[-0.8ex]
\small Paris, France \\ 
\small \texttt{giulia.AJMONEMARSAN@oecd.org}\\      
\and
Nicola Bellomo\\
\small Department of Mathematics, King Abdulaziz University\\[-0.8ex]
\small Jeddah, Saudi Arabia \\ 
\small \texttt{nicola.bellomo@polito.it}\\      
\and
Livio Gibelli\\
\small Department of Mathematical Sciences, Politecnico di Torino \\[-0.8ex]
\small Torino, Italy \\
\small \texttt{livio.gibelli@polito.it}
}

\maketitle

\begin{abstract}
This paper proposes a systems approach to social sciences based on mathematical framework derived from a generalization of the mathematical kinetic
theory and on theoretical tools of game theory.
Social systems are modeled as a living evolutionary ensemble composed by many individuals, who express specific strategies, cooperate, compete
and might aggregate into groups which pursue a common interest.
A critical analysis on the complexity features of social system is developed and a differential structure is derived to provide a general framework
toward modeling.

\noindent
\small {\em Keywords:} Kinetic theory, active particles, stochastic differential games, evolution, learning.

\noindent
\small {\em AMS Subject Classification:} 82D99, 91D10.

\end{abstract}


\section{Motivations and Plan of the Paper}

This essay aims at developing a modeling approach of social agents  viewed as  behavioral,  evolutionary, complex systems. It provides a critical overview on the contribution of mathematical sciences to the modeling, qualitative and computational analysis of social systems with some implications also in the field of economics. Subsequently, a  systems approach is developed and applied to selected case studies.

Various hints coming from a  radical change observed in social and economic sciences~\cite{[ADL97]} suggest to take into account the complexity features belonging to socio-economic systems~\cite{[AR06],[ATV11],[AR06B],[AM09],[ABE08],[ABT13],[BLHKW07],[KIR11],[KZ01],[STIG09]}. The need of including in the modeling approach the specific features of living system is well motivated in the book by Baley~\cite{[BALE90]}, where new concepts of entropy and equilibrium are developed. The same author, in a second book~\cite{[BALE94]} proposes several new ideas toward a theoretical synthesis  for a systems approach to social systems. An additional field of application includes a systems approach to urban planning, as an example papers~\cite{[BAT13],[BT05],[DAC15]}, propose a systems approach of cities viewed as complex systems.

Motivations  are mainly induced by a radical philosophical change that is unfolding in social and economic disciplines. Roughly speaking, the new emerging point of view is characterized by an interplay among Economics, Psychology, and Sociology, which is no longer grounded on the traditional assumption of rational socio-economic behaviors. The rationale for that approach, namely that Economics can be highly affected by individual (rational or irrational) behaviors, reactions, and interactions is widely accepted.

Therefore, an important hallmark in the modeling of socio-economic systems consists in understanding the complexity features of living systems and of
social interactions in particular. Subsequently, mathematical methods need to be developed which are suitable, as far as it is possible,
to capture such features.

The recent literature shows that behavioral features can be taken into account by suitable developments of  methods of statistical mechanics~\cite{[HEL10]}, kinetic theory~\cite{[PT13]}, evolutionary game theory and applications~\cite{[BEIN06],[BEIN12],[CAM03],[GIN09],[HS03],[NOW06],[SPL06],[SVS12],[SBB09]}, mean field games~\cite{[LL07]}, and more specifically by the so-called kinetic theory of active particles~\cite{[BKS13]}, called KTAP, which combines, as we shall see, methods of the classical kinetic theory and theoretical tools of game theory.

Speculations and research activity in the field of physics cannot be neglected by mathematicians, hence the book by Galam~\cite{[GAL13]}, as well as various papers of the same authors such as~\cite{[GAL13A],[GAL13B]}, are  an important reference for the approach proposed in this paper.

Let us now focus with more detail to evolutionary game dynamics which provides an important conceptual contribution to this present paper.
This topic was presented and critically analyzed in the survey paper~\cite{[HS03]}, where the authors put in evidence the substantial difference
compared to the classical game theory~\cite{[NASH1],[NASH2]}.
Namely rather than players involved in a game who attempt to maximize their \textit{payoff}, a whole population is entangled to pursue an
individual or collective wellbeing. This concept is well expressed by the following quotation extracted from~\cite{[HS03]}:

\begin{quote}
\emph{Evolutionary game theory deals with entire population of players, all programmed to use the same strategy (or type of behavior).
Strategies with higher payoff will spread within the population (this can be achieved by learning, by copying or inheriting strategies, or even by
infection). The payoffs depend on the actions of the co-players and hence on the frequencies of the strategies within the population. Since these frequencies change according to the payoffs, this leads to a feedback loop. The dynamics of this feedback loop is the object of evolutionary game theory.}
\end{quote}

Moreover, the authors also observe in~\cite{[HS03]} that one of the great contributions of Mayr  is the replacement of typological thinking by
population thinking~\cite{[MAYR70]}. This concept linked to the dynamics of mutations and selection can explain various aspect of the theory of
evolution~\cite{[MAYR01]}. The authors of this present paper  think that  tools of game theory need including, as we shall see, features of evolution
dynamics, learning processes. An additional important topic, which will be widely treated in the next sections, is the role of nonlinear interactions.

The aforementioned kinetic theory for active particles combines in a general mathematical framework all above concepts including the complex role of
heterogeneous behaviors over the collective dynamics which can include mutations and selection.
Our paper focuses on this approaches and the presentation closely follows the style of a recent book chapter~\cite{[ABHT14]},
where five key questions are posed by the authors to themselves, as well as to the attention of the reader.
The answer to such questions contributes to achieve the aim of developing a  mathematical approach to behavioral social dynamics.

The presentation of this paper accounts for the knowledge that the  research activity in the field is still in a preliminary stage and looks for a commonly shared strategy. Hopefully, different efforts, including those proposed in this paper, might  converge to the common objective of  developing a systems  theory approach.

Bearing all above in mind, the following five key questions are proposed:
\begin{enumerate}
\item {\dc How mathematicians should look at a behavioral social dynamics and to related new research hallmarks?}

\vskip.2cm \item  {\dc Should the derivation of models refer to a general mathematical structure suitable to offer a conceptual framework?}

\vskip.2cm \item  {\dc How far the existing literature has exploited a ``mathematical structure''?}

\vskip.2cm  \item  {\dc Can a general systems theory be developed and how can it contribute to the interpretation of the so-called ``big-data''?}

\vskip.2cm \item  {\dc How and how far, mathematical models can be validated?}
\end{enumerate}

The answer to these  key questions is given in the next sections, where mathematical tools refer to the KTAP's methods reviewed and revisited in~\cite{[BKS13]}. More in detail,  the contents of this paper is guided by  the authors' opinion that suitable  mathematical methods can significantly contribute to a deeper understanding of the relationships between individual behaviors and the collective social outcomes they spontaneously generate. Such mathematical tools should expand the concept of bounded rationality introduced by Simon~\cite{[SIM59],[SIM97]}, according to the idea that choices made by individuals are not perfectly rational as  their rationality is limited by the information they have,
the cognitive limitations of their minds and the time available to make the decision. Decision-makers in this view act as satisfiers who can only seek a satisfactory solution, lacking the ability and resources to arrive at the optimal one, see  also~\cite{[KA12],[TKA74]}.

The alternative approach, which this paper refers to, looks at large social systems whose dynamics is based on individual behaviors, by which single
individuals express, either consciously or unconsciously, their own strategy, which is often based not only on their individual purposes but
also on those they attribute to other agents.

The contents of each section are not limited to technical answers, and a broader analysis  is given looking ahead to perspectives. Adding further indications on the contents:

Section 2 provides, by answering to the first key question, a phenomenological interpretation of social systems which exhibit a behavioral dynamics. The hallmark, which is followed, is  that the modeling approach should be referred to a mathematical structure suitable to retain the complexity features of social systems.

Section 3 focuses on the second key question and derives a general structure which retains the complexity features studied in the preceding section. Such a structure is deemed to provide the framework for the derivation of models.

Section 4 faces the third key question and proposes a critical analysis of the existing literature, also with the aim to
show that some applications fall into the general structure derived in the preceding section.

Section 5  proposes a systems sociology approach, where the mathematical structure can be adapted to the study of a variety of case studies corresponding to different fields of social sciences. Moreover, computational methods and the interpretation of large database are critically analyzed. Hence, this section specifically refers to the fourth key question.

Section 6 presents the analysis of a case study originally developed in~\cite{[BCKS15]} which
models the interaction among citizens, police forces, and criminality. It is shown how various concepts of the preceding sections can be implemented at a practical level.

Section 7 tackles, focusing on the fifth key question, the delicate problem of model validation based on the key problems proposed in the book~\cite{[BL12]}, which will be reported in the next section. Validation cannot be fully achieved for living systems as these are subject to large deviations. However some useful hallmarks can be defined and properly applied.

Section 8 outlines some research perspectives selected according to the authors' bias as one of the aims of this paper consists in addressing future research activity in the challenging field of mathematics interacting with social sciences.


\section{Toward a Behavioral Social Dynamics}
\label{sect:compl.feat}

Although, a model is only an approximation of reality, the contributions of mathematics first to physics and subsequently to applied sciences is well recognized and is in continuous evolution to achieve a deeper understanding of systems of the real world.

Mathematics has developed rather sophisticated qualitative and quantitative tools for studying the inert matter, where causality principles can be generally applied.   On the other hand, the modeling of living matter cannot rely on direct cause-effect links, because the \emph{active} ability of individuals to develop behavioral strategies and to adapt them to the context, makes observable effects come out from often non-evident causes.

The most concrete example is in the field of biology, where despite a long lasting interaction and some successful result, mathematics is still very far from understanding the complexity of biological phenomena. The lack of invariance principles has been well highlighted in biology~\cite{[HER11],[MAY04],[MAYR01]}, while similar arguments can be proposed focusing on the interplay between Mathematics and Socio-Economic Sciences.

This issue is considered in the first key question concerning the conceptual interactions that a mathematician should develop in the
study of social systems.

\vskip.1cm  \begin{center}
{\dc How mathematicians should look at a behavioral social dynamics and to related new research hallmarks?}
\end{center}
\vskip.1cm

Mathematics needs transferring what is observed through a phenomenological interpretation of social reality into equations, generally of differential nature. Therefore it is important selecting a number of key features toward the aforesaid mathematical purpose.

In general, living systems exhibit various complexity features. In particular, interactions among individuals need not have an additive linear character. As a consequence, the global impact of a given number of entities (\emph{field entities}) over a single one (\emph{test entity}) cannot be assumed to consist merely in the linear superposition of the actions exerted individually by single field entities. This nonlinear feature represents a conceptual difficulty to the derivation of mathematical models. More generally, the new point of view presents social science as a discipline, which should deal with  evolving complex system, where interactions among heterogeneous individuals and the interplay among different dynamics can  even produce unpredictable emerging outcomes~\cite{[ADL97],[BHT13]}.

After these preliminary statements, the following complexity features are selected:

\begin{itemize}
\vskip-.2cm\item {\dc Ability to express a strategy:} \textit{Living entities are able to develop specific strategies to fulfill their goals depending on their own state and on that of the entities in the surrounding environment. Living systems typically operate to achieve their well-being. The said strategies are not defined once for all, but are modified according to the changing environment where individuals operate.}

\vskip.2cm \item {\dc Heterogeneity:}  \textit{The ability to express a strategy  is heterogeneously distributed. In the case of social (and economic) systems, heterogeneous behaviors can play an important role in determining the overall collective dynamics~\cite{[AR06]}, whenever irrational behaviors of a few entities can induce large deviations from the usual dynamics observed in rationality-driven situations.}

\vskip.2cm \item {\dc Nonlinear interactions:}  \textit{Interactions are nonlinearly additive and involve immediate neighbors, but in some cases also distant individuals due to the ability living systems to communicate. In some cases, the topological distribution of a fixed number of neighbors can play a prominent role in the development of strategies and interactions.
Self organization is an important consequence of interactions.}

\vskip.2cm \item {\dc Learning and adaptation:}  \textit{Living systems have the ability of learning from past experience. Therefore, their strategic ability and the characteristics of mutual interactions evolve in time due to inputs received from outside. As a result, continuous adaptation to the changing-in-time environmental conditions occurs.}

\vskip.2cm \item {\dc Selection and evolution:}  \textit{Aggregation of individuals can be viewed, as the formation of groups sharing the same interest~\cite{[ABT13]}, which may generate new groups more suited to an  evolving social and economical environment as well as groups less suited  to the social context. In the course of time, some of such  groups may  disappear, and new ones may become dominant.}

\end{itemize}

According to these features, it is apparent that the dynamics of a few entities does not lead straightforwardly to that of the whole system,
because the latter manifests, at a collective level, the nonlinear effects of individual interactions.
This observation can be further understood based on the following statements, extracted from~\cite{[ABT13]}:

\begin{quote}
\emph{Living entities express, rationally or irrationally, a certain strategy for achieving their own wellbeing. These systems are often complex:  At large scale collective behaviors appear, which are apparently coordinated but are actually \emph{self-organized}. Namely, they emerge spontaneously without the action of any external organizing principle.}
\end{quote}

\begin{quote}
 \emph{Sudden deviations from the usual standards may occur, leading to even highly unpredictable events with dramatic collective consequences, as well demonstrated nowadays by recent facts in our societies. One of such events, which is of paramount interest in Social Sciences in general, and in Economics in particular, is the so-called \emph{Black Swan}, namely an extreme event, largely unpredictable at a collective level, originating from apparently rational and controlled individual behaviors~\cite{[TAL07]}.}
\end{quote}

\begin{Remark}
 Complexity in living systems is induced by a large number of variables, which are needed to describe their overall state. Therefore, the number of
 equations needed in the modeling approach might be too large to be practically treated.
 Appropriate methods need thus to be developed with the aim of reducing this structural complexity.
 At the same time these methods should be multiscale in nature since using only a single observation and representation scale is not sufficient to
 describe the dynamics of living systems. For instance, the micro-scale of individuals affects the scale corresponding to a group of interest,
 which in turn has a direct influence on the dynamics of large systems made up of the union several interacting groups of interest.
\end{Remark}

\begin{Remark}
Living systems composed by many individuals show collective emerging behaviors that are not directly related to the dynamics of a few entities, but are often generated by a kind of swarming intelligence that involves all the interacting individuals~\cite{[BS12],[BDT99],[COU07]}. At a qualitative level, emerging behaviors are often reproduced  under suitable  input conditions, though quantitative matches with the observations are rarely obtained. In fact, small changes in the input conditions  often generate large deviations. In some cases, these break out the macroscopic (qualitative) characteristics of the dynamics, whence substantial modifications can be observed. Heterogeneity of individual strategies, learning ability, and interactions with the outer environment largely influence such phenomena. According to~\cite{[BHT13]}, these deviations can be interpreted as the requisites for a ``Black Swan'' to develop~\cite{[TAL07]}. More precisely, the following definition can be extracted from the cited book:

\begin{quote}
\emph{A Black Swan is a highly improbable  event with three principal characteristics: it is unpredictable; it carries a massive impact; and, after the fact, we concoct an explanation that makes it appear less random, and more predictable, than it was.}
\end{quote}
\end{Remark}

\begin{Remark}
Mathematical models can hopefully offer a powerful tool to show the evolution of different possible scenarios in a socio-economic systems. They might even lead to better understand what possible directions emerging collective actions may take,  depending on different initial conditions and individual behaviors.
\end{Remark}

All the remarks above have been presented in agreement with the authors' bias. However, it is not claimed that this section offers an exhaustive framework. Therefore, we wish to present the  key problems proposed by Bonacich and Lu~\cite{[BL12]}, with the aim of getting back to them for further speculations after having presented the mathematical tools proposed by our paper.

\begin{enumerate}

\item \textit{Regarding networks and homophyly birds of feather flock together, but people are influenced by those they like. Both these processes result in the same outcome, but there is no standard accepted way of separating these  two processes.}

\vskip.1cm \item  \textit{There are lots of models that show how groups arrive at consensus but no generally accepted model of how groups become more and more different and possibly hostile. Moreover,  what are the most important mechanism for bringing about cooperations in groups?}

\vskip.1cm \item  \textit{We know that people are affected by their positions in networks, but we do not have a variety of models of how people create their networks. We also do not have good models for network change and evolution.}

\vskip.1cm \item  \textit{There are a variety of measures of centrality in networks, but there are no well-established criteria for when one measure is preferable to another.}

\vskip.1cm \item  \textit{Are human groups unpredictable because humans themselves are complex organisms or because there are truly chaotic dynamics in groups?}

\end{enumerate}

We have reduced  the  ``key problems'' from the six proposed in~\cite{[BL12]} to five, simply by grouping the previous second and fourth problem. Actually, the authors refer in~\cite{[BL12]} to the celebrated Hilbert's problems, which have highly engaged the mathematical research activity in the past years and are still
object of challenging speculations. Although the reference to Hilbert appears to us somehow excessive, still we think that these problems offer possible way  to validate models and theories or, at least, critically analyze them. Therefore, the whole matter will be treated again in Section 5.


\section{On the Search of a Mathematical Structure}
\label{sec:structure}

This section provides an answer to the second key question, which focuses on the search of structure suitable to capture the specific features of social systems proposed in Section 2.
\vskip.1cm
\begin{center}
{\dc Should the derivation of models refer to a general  mathematical  structure  suitable to offer a conceptual framework?}
\end{center}
\vskip.1cm

The answer is given through four subsections.
Subsection 3.1 provides a general description of a large class of social systems object of the modeling approach and outlines the modeling
strategy which generates the mathematical structure we look for.
Subsection 3.2 defines the representation of the system by
linking at each functional subsystem a probability distribution over the microscopic state.
In subsection 3.3 the aforementioned structure is derived.
Finally, some reasonings on the modeling of interactions are discussed.


\subsection{An heuristic description of social systems}

 Let us now define precisely the class of systems which is object of the modeling approach. In detail, we consider a network of a number $m$ of interacting nodes, where in each node  individual entities are  grouped into $n$ functional subsystems.

According to~\cite{[BKS13]}, the  hallmarks of the modeling approach  can be summarized as follows:

\begin{itemize}

\item The entities which comprises the system are referred to as \emph{active particles} and are assumed to be distributed in a network of nodes labeled by the subscript  $i=1, \ldots, m$. Therefore, the role of space variable is confined  to the network, namely a continuous variation in space is replaced by migration dynamics across the nodes.

 \vskip.2cm  \item Active particles are  partitioned into \emph{functional subsystems}, where they  collectively develop  a common strategy. These are labeled by the subscript $j= 1, \ldots, n$. Therefore the total number of subsystems is $N= m\, n$.

 \vskip.2cm  \item The strategy is heterogeneously distributed among active particles and corresponds to an individual state, defined \emph{activity}. Hence the state of each functional subsystem is defined by a probability distribution over the said activity variable.

\vskip.2cm \item Active particles interact within the same functional subsystem as well as with particles of other subsystems, and are acted upon the external actions. These interactions are generally nonlinearly additive and are modeled as \emph{stochastic games}, meaning that the outcome of a single interaction event can be known only in probability.

\vskip.2cm \item The evolution of the probability distribution is obtained by a balance of particles within elementary volumes of the space of microscopic states, the inflow and outflow of particles being related to the aforementioned interactions.
\end{itemize}

These hallmarks are quite general and need to be more precisely defined depending on the system under consideration.
In particular, the following aspects have to be accounted for:

\vskip.1cm \no i) Definition of the functional subsystems and modelling of their dynamics across nodes;

\vskip.1cm \no ii) Modelling of the transitions across functional subsystems;

\vskip.1cm \no iii) Presence of leaders, which should be grouped in specific functional subsystems;

\vskip.1cm \no iv) Characterization of the so-called external actions and of the dynamics induced by them;

\begin{Remark}
A vast literature has been recently developed on the theory of networks and applications,
for example~\cite{[ABO10],[AB02],[BFP09],[BAR12],[BBV06],[EMV06],[GL05],[GL08A],[RAC11],[VR07]}.
This present paper refers only tangentially to such a theory, while some theoretical tools of evolutive game theory are used to model interactions also across the network.
\end{Remark}


\subsection{Functional subsystems and their stochastic representation}

The overall state of the internal system is delivered by a probability distribution over  the  activity variable in each node and for  each functional subsystem, hereinafter denoted by the acronym $ij$-FS, while the generic one is simply denoted by FS.
In detail, the probability distribution reads
\begin{equation}\label{probability}
f_{ij}(t, u)\,: \quad [0,T] \times D_u\, \to\, \RR_+,
\end{equation}
such that moments can provide a description of quantities of the system at the macroscopic scale. In detail the zeroth order moment gives the number density:
\begin{equation}\label{moments_low}
 n_{ij} [f_{ij}] (t) = \E_{ij}^0 [f_{ij}](t) = \int_{D_u} f_{ij}(t, u)\,du,
\end{equation}
which gives the number of active particles in each  $ij$-FS. Higher order moments, corresponding to additional macroscopic variables, can be computed as follows;
\begin{equation}\label{moments_high}
\E_{ij}^p(t) [f_{ij}] = \frac{1}{n_{ij} [f_{ij}] (t)}\, \int_{D_u} u^p f_{ij}(t, u)\,du, \qquad p = 1, 2, \ldots,
\end{equation}
where  the  computing of macroscopic quantities is meaningful  only if the number of active particles in the FS is a strictly positive defined quantity.

 The activity is represented by scalar variable whose domain of definition $D_u$   is typically $[-1,1]$ or even the whole real axis for probability distributions which decay to zero at infinity. Positive values indicate an expression coherent with the activity variable, while negative values expressions in opposition.  The interval $[-1,1]$ is used whenever a maximum admissible value of the activity variable can be identified so that $u$ is normalized to such value.

 More in general the activity variable can be a vector, but we remain here within the relatively simpler case of a scalar variable, which makes light the formalism of the mathematical approach.

 The actions exerted by the \textit{external system} are supposed to act at the same scale of the system and over the same domain of the activity variable by agent systems. These actions are assumed to be known functions of time and of the activity variable defined by the following structure:
\begin{equation}\label{external}
\vf_{ij}(t,u) = \ve_{ij}(t)\, \psi_{ij}(u)\,: \quad [0,T] \times D_u\, \to\, \RR_+,
\end{equation}
where $\ve_{ij}$ models the intensity of the action in the time interval $[0,T]$, while $\psi_{ij}$ models the intensity of the action over the activity variable.


\subsection{Mathematical structures and models of interactions}

The mathematical theory of active particles~\cite{[BKS13]} leads to the derivation of mathematical structures, which describe the evolution in time of the probability
distribution over the activity variable for each functional subsystems. The result, which  consists in a system of   $n \times m$  differential-integral equations, is obtained by a balance of particles within the elementary volume  $[u, u + du]$ of the space of the microscopic states. This approach is documented in the already cited survey~\cite{[BKS13]} as well as in various papers concerning specific applications such as the immune competition \cite{[BE15],[DEA14]} or learning dynamics~\cite{[BDG15]}. Therefore, only a concise description of the technical aspects of the approach will be given here referring  to the existing literature for further details.

The dynamics of interactions can be obtained by assigning to each particles of the internal $ij$-FS the  following distinguished roles: \textit{Test particle} of the $i$th node and $j$th FS with activity $u$, at time $t$; \textit{Candidate particle} of the $i$th node and $j$th FS with activity $u_*$, at time $t$, which  can gain the test state $u$ as a consequence of the interactions; \textit{Field particle}  with activity $u^*$, at time $t$, which  triggers the interactions of the candidate particles. Hereinafter, the term \textit{$(ij)$-particle} is used to denote a particle in the $ij$th FS. It is assumed that the overall state of the system is delivered by the probability distribution linked to the test particle.

Moreover, the following interactions occurring at different scales are considered:
\begin{itemize}
\item  \textit{{\dc Microscopic scale interactions}: Individual based interactions between particles belonging to the same FS or to different FSs.}
\vskip.2cm \item \textit{{\dc Microscopic-macroscopic scale interactions}: Interactions between particles and FSs viewed as a whole
being represented by their mean value.}
\end{itemize}

The mathematical framework describing different kind of interactions will be specified, for each type of interactions, by means of two terms:
\begin{itemize}
\item The \textit{interaction rate} which describes the rate of interactions involving  particles;

\vskip.2truecm \item The \textit{transition probability density} which describes the probability density that a candidate particle falls into the state of the test particle
after the interaction.
\end{itemize}

Details on the modeling of these  interaction terms will be given in the next subsection. Supposing now that these terms can be defined by appropriate models, the following balance of number of particles can be used to derive the mathematical structure:
\begin{align*}
&\textit{Dynamics of the number of active particles} \\
&\quad  \textit{ = Net flux due to ms-interactions between active particles} \\
&\qquad \textit{+ ms-interactions between active particles and external actions}  \\
& \quad \qquad \textit{+ mMs-interactions  between particles and functional subsystems},
\end{align*}
where the net flux denotes the difference between inlet and outlet fluxes, while ``ms'' stands for microscopic scale and ``mMs'' stands for microscopic-macroscopic scale.

Detailed calculations are not reported here as the interested reader can find them in some recent papers such as~\cite{[BDK13],[DEA14]} and \cite{[BE15]}. The result of the said calculations is the following structure:
 \begin{equation} \label{eq:structure}
\partial_t f_{ij}(t,u) = A_{ij}[\f](t,u) +  B_{ij}[\f,\vf](t,u) + C_{ij}[\f](t,u), \\
\end{equation}
where
\begin{eqnarray}\label{I}
 A_{ij}[\f](t,u) &=& \sum_{p,h=1}^m \sum_{q,k=1}^n  \iint_{D_u\times D_u}\hspace{-.3truecm}
\eta_{pq}^{hk}[\f](u_*,u^*)\, \mathcal{A}_{pq}^{hk}(ij)[\f](u_* \to u|u_*,u^*) \nonumber \\
&\times& f_{pq}(t,u_*)\,f_{hk}(t,u^*)  \,du_*\,du^*\nonumber \\
 &-&  f_{ij}(t,u) \sum_{h=1}^m\sum_{k=1}^n  \int_{D_u} \eta_{ij}^{hk}[\f](u,u^*)\,f_{hk}(t,u^*)\, du^*,
\end{eqnarray}
\begin{eqnarray}\label{II}
 B_{ij}[\f](t,u) &=& \sum_{p=1}^m \sum_{q=1}^n  \iint_{D_u\times D_u}\hspace{-.3truecm}
\mu_{pq} [\f,\vf](u_*,v^*)\, \mathcal{B}_{pq}(ij)[\f,\vf](u_* \to u|u_*,v^*) \nonumber \\
&\times& f_{pq}(t,u_*)\,\vf_{pq}(t,u^*)  \,du_*\,du^*\nonumber \\
 &-&  f_{ij}(t,u)   \int_{D_u} \mu_{ij}[\f,\vf](u,v^*)\,\vf_{ij}(t,u^*)\, du^*,
\end{eqnarray}
\begin{eqnarray}\label{III}
 C_{ij}[\f](t,u) &=& \sum_{p,h=1}^m \sum_{q,k=1}^n  \int_{D_u}
\nu_{pq}^{hk}[\f](u_*)\, \mathcal{C}_{pq}^{hk}(ij)[\f](u_* \to u|u_*,\E_{hk}[\f]) f_{pq}(t,u_*) \nonumber \\
&\times& \E_{hk}[\f]\,du_* \nonumber \\
 &-&  f_{ij}(t,u) \sum_{h=1}^m \sum_{k=1}^n  \nu_{ij}^{hk}[\f](u)\,\E_{hk}.
\end{eqnarray}

Moreover, the \textit{interaction rates}, which model the frequency of interactions  for each type of interaction have been denoted by $\eta_{pq}^{hk}, \mu_{pq}$ and  $\nu_{pq}^{hk}$. The \textit{transition probability densities}, which denote the probability density that a candidate $pq$-particle shifts, after an interaction with a field $hk$-particle,
to the state of the test $ij$-particle  have been denoted by $\mathcal{A}_{pq}^{hk}, \mathcal{B}_{pq}$, and $\mathcal{C}_{pq}^{hk}$. The output of interactions has been put into round brackets. Square brackets have been used to denote dependence on the distribution functions (nonlinear interactions).

\begin{Remark}
It has been assumed that each node includes all FSs. This assumption is formal if the initial status of some FSs is equal to zero and remains like that for $t >0$. On the other hand, such status  can take values for $t > 0$  due to migration phenomena.
\end{Remark}

\begin{Remark}
This structure includes some specific systems. As an example, the case of a closed system  can be considered by setting $\vf = 0$. The structure simplifies as follows:
 \begin{equation} \label{structure-s_one}
\partial_t f_{ij}(t,u) = A_{ij}[\f](t,u)  + C_{ij}[\f](t,u),
\end{equation}
On the other hand, a possible alternative  to the structure can be obtained by modeling an external action $K_{ij}(t,u)$ acting over each functional subsystem, which yields:
 \begin{equation} \label{structure-s_two}
\partial_t f_{ij}(t,u) + \partial_u \big[K_{ij}(t,u) f_{ij}(t,u)\big] = A_{ij}[\f](t,u)  + C_{ij}[\f](t,u).
\end{equation}
Or by supposing that the external actions acts over the system by the mean value of $\psi_{ij}$ so that the term $B_{ij}$ is modified accordingly. Further generalizations can be achieved by taking into account also macroscale actions by each node as a whole and by the whole network.
\end{Remark}

\begin{Remark}
The output of interactions between two interacting entities can be conditioned by their states at the microscopic scale and by the probability distribution over these states. As previously mentioned, interactions are  \textit{linear} if the output is conditioned only by the microscopic states, while are  \textit{nonlinear} when the output is conditioned by the probability distributions or by their moments.  The various sources of nonlinearity featured by the equation above are critically analyzed in~\cite{[BC13]}.
\end{Remark}

\begin{Remark}
The term \textit{stochastic games} is used since the state of the interacting entities, as well as the output of their interactions, are known in
probability. Moreover, the term  \textit{evolutionary} is used to account for the dynamics in time of the rules that drive interactions. Since these games are casted into a differential system, we will talk about  \textit{stochastic evolutionary differential games}.
\end{Remark}

\begin{Remark}
Several applications, as examples~\cite{[ABT13],[BHT13]}, suggest to use discrete variables at the microscopic scale. In fact, in some specific cases the state of the active particles is more precisely identified by means of ranges of values rather than by a continuous variable.
Technically, this means that the distribution functions, for each FS, have to be regarded as discrete distribution functions, while the interaction terms map discrete variables rather than continuous ones. The mathematical structures are readily obtained by specializing the integrals over the
microscopic states as sums over discrete states. We skip over the technical, it might be tedious, derivation of a general expression of a mathematical framework for systems with discrete
states, as it is more practical focusing such a problem on specific models rather than on a general case. This topic will be treated, at a practical level, later in Subsection 5.2 when the general approach will be focused on a case study.
\end{Remark}


\subsection{Some reasonings towards modeling interactions}

Specific models of social dynamics can be obtained by inserting  models suitable to depict interactions at the microscopic scale
into the structure~\eqref{eq:structure}. Various recent papers have contributed to this topic. For instance~\cite{[BCKS15],[BHT13],[DL14]} have shown how interactions
can be modeled by games where the output of the interactions is conditioned not only by the state of the interacting entities, but also by the probability distribution over such states.
 Moreover, models of multiscale micro-macro interactions have been proposed in~\cite{[KNO13]} concerning the modeling of migration phenomena, in~\cite{[KNO14]} on opinion formation in small networks,
   and in~\cite{[BCKS15]} focused on criminality dynamics. These multiscale features  deserve attention in several fields of social dynamics.

The aforementioned literature suggests to critically revised the whole subject.
Such a revisiting is needed to derive tools that are valid for a large variety of social systems, as well as to deal with
the actions externally exerted on the system, which is a topic only tangentially treated in the literature although very important in the modeling of social systems.

Bearing all above in mind, let us consider, separately, the modeling of the interaction rates and of the transition probability density.

\subsubsection{Interaction rate}

The modeling of interactions requires the definition of different concepts of \textit{distance} between
interacting entities~\cite{[BCKS15]}. In details:
\begin{itemize}
\item \textit{Microstate-microstate distance} refers to the states at the microscopic scale  between  candidate and  field particles. This distance can be $|u_*-u^*|$.

\vskip.2truecm \item \textit{Microstate-macrostate distance} refers to the interaction of a candidate particle with state $u_*$ with the group of particles belonging to a FS with mean value of the activity $\E$.
      This distance is given by $|u_*-\E|$.

\vskip.2truecm\item \textit{Affinity distance} refers to the interaction between active particles characterized by different distribution functions.  This distance is introduced according to the general idea that two systems with close distributions are \textit{affine}   and is given by $||f_{pq}-f_{hk}||$, where $||\cdot||$ is a suitable norm to be chosen depending on the physics of the system  under consideration.

\vskip.2truecm\item \textit{Hierarchy distance} whenever a hierarchy exists among groups of interest to distinguish the more and less relevant ones.

\end{itemize}

The overall distance, let us call it \textit{social metrics}, is a weighted sum, which replaces the geometrical metrics, of all distances.
In general, the interaction decays with the distance, where heuristic assumptions lead to a decay described by exponential terms or rational fractions.

\subsubsection{Transition probability density}

The dynamics of interactions can be modeled by theoretical tools from evolutionary and behavioral game theory \cite{[CAM03],[GIN09],[HEL10],[NOW06],[SPL06],[SVS12],[SBB09]}, which provides features to be introduced into the general mathematical structure in order to obtain specific models. Additional tools are given by learning theory~\cite{[BDG15]} and evolutionary games~\cite{[NOW06]}, within the broader context of statistical dynamics and probability theory. For instance, the following types of games are often used:

\begin{itemize}
\item \textbf{Competition (dissent)}: The interacting particle with higher status increases its status by taking advantage of the other with lower status. Therefore the competition is advantageous for only one of the two players involved in the game.

\item \vskip.2truecm\textbf{Cooperation (consensus)}: The interacting particle shows a trend to share their microscopic states by decreasing of the difference between the interacting particles' states, due to a sort of attraction effect~\cite{[HY09],[MT14]}.

\item \vskip.2truecm\textbf{Learning}: One of the two particles  modifies, independently of the other, its microscopic state, while the other reduces the distance by a learning process.

 \item \vskip.2truecm\textbf{Hiding/chasing}: One of the two particles  attempts to increase the distance from the state of the other one (\emph{hiding}), which conversely tries to reduce it  (\emph{chasing})~\cite{[BCKS15]}.

     \end{itemize}

\begin{Remark}
 All aforesaid types of games can occur simultaneously in a general context of heterogeneous particles. In some cases, see~\cite{[BHT13]}, the occurrence of one of them is ruled by a \emph{threshold} on the distance between the states of the interacting particles. Such a threshold is, in the simplest case, a constant value. However, it is worth mentioning that a recent paper~\cite{[DL14]} has shown that the threshold can depend on the state of the system as a whole and that can have an important influence on the overall dynamics. Thus dynamics  refers to the interaction of more than one type of dynamics as treated in \cite{[BPS10]} from the view point of social economy, where political choices are related to economical growth. Further applications, in addition to the already cited paper  \cite{[DL14]}, are proposed in~\cite{[BHT13]} and~\cite{[BCKS15]}.
\end{Remark}

 Let us now consider the above remark referring more in detail~\cite{[DL14]}, where a parameter has been introduced to account for the ``selfishness'' of the society. Such parameter is related to the evolution equation describing the dynamics of the aforementioned threshold. A detailed study of the model shows that high values of the selfishness in the wealth distribution ends up with an overall reduction of the total wealth, which is shared only by small groups, while the great part of the population is confined in poverty.  The mathematical structure has been modified in~\cite{[DL14]} by linking the equation for the dynamics of the distribution function to an equation for the threshold triggering interactions. The formal structure in the general case is as follows:
\begin{equation}
\label{2.2}
\begin{cases}
\partial_t f_{ij}(t,u) = A_{ij}[\f,\mu](t,u)  + C_{ij}[\f,\mu](t,u),\\
\\
\partial_t \mu_{ij} = F(\mu, \f),
\end{cases}
\end{equation}
where $\mu$ denotes the set of all thresholds $\mu_{ij}$. A deep insight into the role of selfishness and the complex problem of its ``measurement'' is proposed in~\cite{[SIG11]}. The effect of network topology on wealth distributions is studied in~\cite{[GL08B]}.

Based on simulations, Paper~\cite{[DL14]} predicts that in an excessively selfish society  wealthy classes initially enrich themselves
but eventually the whole society moves toward poverty since the middle classes, which generally are the producers
of positive growth of the market, are strongly weakened.

\begin{Remark}
The interaction domain $D_u$ has been supposed, for simplicity, constant. However, it can be assumed to be depending on the distribution function.
 A possible example is that of~\cite{[BS12]} based on the conjecture that sensitivity is related to a critical size rather than to the whole perception domain.
  This concept deserves attention to be related to specific models.
This remark accounts for~\cite{[BS12]},  where it is suggested that interactions occur in a domain a sensitivity domain $\Omega \subseteq D_u$,
corresponding to a critical number $n_c$ of field particles. Integration of $f_i$ over the activity variable in a domain  $\Omega = [u - s_m[f;n_c], u + s_M[f;n_c]]$,
which can be called  the topological domain of interaction, can lead to compute $s_m$ and $s_M$, where $s_m, s_M > 0$. However, the solution is unique only in some special cases.
For instance, when $u$ is a scalar defined over the whole real axis, and the sensitivity is symmetric with respect to $u$.
On the other hand, if $u$ is defined in a bounded domain or the sensitivity is not symmetric, additional assumptions are required.
\end{Remark}


\section{A Critical Overview of the Existing Literature}
\label{literature}

The mathematical structure proposed in the preceding section has been already used, simplified to special cases, by some models known in the literature. Therefore, it is worth developing an overview of these particular models to understand which type of innovation can be achieved by a more general approach. This analysis can contribute to show how the various approaches are related to the complexity features of social systems.  These two topics are presented in the next two subsections as an answer to the third key question:
\vskip.1cm
\begin{center}
{\dc How far the existing literature has exploited a ``mathematical structure''?}
\end{center}
\vskip.1cm

Before presenting a survey of the existing literature, let us mention that the search of a structure, which is here viewed as a necessary step of the modeling approach, can have a broader meaning in various fields of life sciences as observed by Gromov~\cite{[GRO12]}, who stresses that a deep analysis  of mathematical structures might offer an overview much reacher than that foreseen by the initial application. Referring to the specific structure under consideration, suitable developments such as the introduction of proliferative and destructive interactions have generated models in biology~\cite{[BDK13]}, while the introduction of a space structure have lead to models of vehicular traffic and crowds~\cite{[BBNS14],[BDF12],[BB15b]}.


\subsection{An overview of models}
The application of kinetic type equations with interactions modeled by theoretical tools of game theory was introduced in~\cite{[ABL00]} and subsequently paper~\cite{[BD04]} generalized such a structure to model large social systems with discrete states  at the microscopic scale. The authors used games with linear binary interactions and derived a system of ordinary differential equations suitable to describe the time dynamics of a discrete  probability distribution corresponding to the discrete micro-states. Actually, the authors used a simplified version of a framework proposed in~\cite{[BDEL06]} to model complex multicellular systems. The application developed in~\cite{[BD04]} refers to the dynamics of wealth distribution and is based on consensus/dissent games triggered by a constant threshold of the distance between the states of the interaction pairs.

The same approach was subsequently extended with minor technical modifications to various types of applications by the authors and their coworkers.
Without claim of completeness, examples include opinion formation~\cite{[BD08]}, taxation dynamics with some social implications~\cite{[BM11]}, value estimates~\cite{[DL13]}. Various innovative developments followed these pioneer papers. The main sequential steps of the aforementioned developments are summarized in the following looking ahead to new research approaches,
which can be achieved by use of the structure of Eq.~(\ref{eq:structure}).

Opinion formation is an important ground of application of mathematics to the modeling of social sciences, hence several valuable papers have been
devoted to this topic, for example~\cite{[ANT07],[BST09],[DMP12],[HK02],[HK15],[TOS06]} have developed interesting models based on kinetic theory methods. These methods have also been applied to modeling financial markets~\cite{[CPT05],[CPP09]}.

\sm  \textbf{A pioneer approach to systems theory:} Various authors have proposed different approaches to a systems theory of social dynamics, for instance \cite{[BALE94],[BL12]}. However, the first contribution along the conceptual line of this present paper  has been given in~\cite{[ABE08]}, where the concept of scaling and functional subsystems have been proposed referring to a variety of social systems as well as the observation that modeling the interactions of different types of dynamics is needed. This is, indeed, a topic of great interest in socio-political sciences~\cite{[BPS10]}.

\sm \textbf{On the introduction of nonlinear microscopic interactions:} Nonlinearity for interactions at the micro-scale have been introduced in \cite{[BHT13]} and~\cite{[DL14]}. Both  papers analyze the interplay of different dynamics, namely wealth distribution versus support or opposition to Governments, or selfishness in wealth dynamics as a source of overall wealth dissipation in nations. The role of nonlinearity of interactions was clearly put in evidence in~\cite{[BC13]}.
An interesting reference for social aspects of selfishness and wealth distribution is given by \cite{[PSC14]}, based on previous studies in the field~\cite{[GP09],[KPK09],[KPM12]}.

\sm  \textbf{Dynamics over small networks:} The modeling of this type of dynamics by kinetic theory methods was introduced in~\cite{[KNO13]} and~\cite{[KNO14]} referred to specific applications, such as migration phenomena. These two papers have introduced a new type of nonlinearity under the assumption that individuals are sensitive
to the mean value of each functional subsystem, which depends on the distribution function. Indeed, this is the two scales micro-macro interaction presented in our paper.
An alternative approach to modeling migration dynamics are presented in~\cite{[BO12]}  and~\cite{[BOR89]}. The concept of social distance, introduced in Subsection 4.3.1,
 is one of the main tools toward the study of networks. Such a distance can be viewed as a stochastic quantity as it depends on a probability distribution.
  Therefore the  structure of a social network evolves in time and is not rapidly related to  the physical localization.

\sm  \textbf{Mutations and selection:} A post Darwinian dynamics consisting in mutations followed by selection  plays an important role in biology, specifically in the immune competition~\cite{[BE15],[BDK13],[DEA14]}. An analogous dynamics appears in socio-economical systems, where new groups of interest can be generated in the dynamics, for instance by aggregation of different groups, and subsequently they might either expand or disappear in a competition somewhat mediated by the environment where these groups act. These concepts have been introduced in some recent papers, for instance~\cite{[DAC15]} in a behavioral theory of urbanism.

\sm  \textbf{Control problems:} An important field of investigation  is the development of a control theory of social systems. This topic has been developed in the study of swarm dynamics \cite{[BW15],[CFPT15]} and recently focused more precisely on the dynamics of social systems~\cite{[CLP15]}. The mathematical structures proposed in this paper include the presence of external actions to be properly studied to drive the dynamics toward optimal behaviors.


\subsection{Descriptive ability of the mathematical structures}
A critical analysis of the capability of the mathematical structure (\ref{eq:structure}) to capture the complexity features presented in Section 2 is a necessary step toward the systems approach.  In detail:

\begin{itemize}
\vskip-.2cm\item {\dc Ability to express a strategy:} \textit{ The microscopic state includes the activity variable, which is deemed to model the strategy expressed by each functional subsystem.}

\vskip.2cm \item {\dc Heterogeneity:}  \textit{The heterogeneous behavior is accounted for by the description of the system by means of a probability distribution  linked to each functional subsystem.}

\vskip.2cm \item {\dc Nonlinear interactions:}  \textit{Nonlinearity of interactions has been already discussed in the preceding section. As a result linear additivity cannot be applied.  Moreover, we observe that an additional source of nonlinearity is related to  non-locality of interactions between functional subsystems localized in different nodes of the network.}

\vskip.2cm \item {\dc Learning and adaptation:}  \textit{The ability of living systems of learning from past experience can be mimic by  modeling  interactions  based on rules that evolve in time and include a continuous adaptation to the changing-in-time environmental conditions.}

\vskip.2cm \item {\dc Selection and evolution:}  \textit{The onset of functional subsystems is modeled by the transition probability density, where interactions include the formation of groups of interest, which in turn can generate new groups more suited to an  evolving social and economical environment. Selection is modeled by interactions with the outer environment.}
\end{itemize}

It is worth noticing that the descriptive ability of the complexity features is only potential and it should be practically implemented in the derivation of models when theoretical tools of game theory are used to describe interactions.
Although models rapidly reviewed in Subsection 4.1, have been successfully depicted some collective behaviors observed in reality,
it can be remarked that one of the limits of the existing literature is that it deals with closed systems, while interactions with the outer environment has not been taken into account.
Moreover, the role of networks in determining the overall dynamics has been treated only tangentially. Therefore, a great deal of work is still waiting for being developed. Hopefully, the next sections will contribute to link the conceptual ideas proposed until now to real applications.


\section{Toward a Systems Approach to Behavioral Social Dynamics}
\label{systems}

This section shows how the mathematical tools presented in Section 3 can be applied to modeling and simulation of social systems. The aim consists in providing an answer to the fourth key question:
\vskip.1cm  \begin{center}
 {\dc Can a general systems theory be developed  and how can it contribute to the interpretation of the so-called ``big-data''?}
\end{center}
\vskip.1cm

With this aim in mind the contents of this section can be presented. Subsection 5.1 provides the general hallmarks of the theoretical approach. Then, Subsection 5.2 provides a critical overview on Monte Carlo  computational methods, which appear to be the most efficient method for the class of equations proposed in this paper.  Finally, Subsection 5.3 shows how the system approach can contribute to the use and interpretation  of repositories of the so-called big data toward the interpretation of the dynamics of social systems. This section will remain at a theoretical level, while a specific case study is presented in Section 7, where various concepts are referred to a well defined model.


\subsection{Hallmarks of a systems theory}

Some guidelines towards a systems approach have already been anticipated in Subsection 3.1, however not yet referred to any mathematical formalization. The mathematical tools derived in Section 4 allow us to proceed in this crucial step. Therefore, a revisiting of the previous concepts is proposed as follows:

\begin{enumerate}

 \item The domain  of the space variable where the system is located is subdivided into interconnected sub-domains called \textit{nodes},
 while the set of all nodes and their connection is called  \textit{network}.

\vskip.2cm \item  The individual entities, called \emph{active particles}, are aggregated into different groups of interest called
 \emph{functional subsystems}.

\vskip.2cm \item  The active particles aggregated in a functional subsystem express a common strategy (or expression of interest)
 called \textit{activity}.

\vskip.2cm \item The heterogeneity of the particles' activity within each functional subsystem is expressed through a
 \textit{probability distribution}. The overall set of probability distributions represents the dependent variable of the dynamics.

\vskip.2cm \item  Modeling is required for interactions between active particles  within the same functional subsystem and with particles of other
subsystems, as well as for \textit{external actions}.  All these interactions can induce transitions across functional subsystems.

\vskip.2cm \item The equations modeling the dynamics of the  probability distributions over the micro-states are obtained by a balance of particles within elementary volumes of the space of microscopic states, the inflow and outflow of particles being related to the aforementioned interactions.

\vskip.2cm \item These equations can be solved by suitable computational methods to obtain both the probability distributions
 and the moments which provide an information of interest on macroscopic states.

\end{enumerate}

\begin{Remark}
The following specific milestones of the approach: \textit{Subdivision, activity variables, external actions, network, etc.} are not  constant features of the overall system. In fact, these variables depend on the specific social dynamics object of study. Therefore, different investigations would lead different characterization of the overall system and hence of mathematical models.
\end{Remark}


\subsection{Computational methods}

The development of the numerical methods for solving the class of equations proposed in this paper should take into account that the
dependent variables are probability distributions over the activity variable depending on time.
Deterministic solutions can be achieved by approximating the said distribution at a number of collocation points over the activity variable.
This transforms the original integro-differential system into a coupled system of ordinary differential equations corresponding to the discrete
values of the probability distributions in each point of the collocation.
Classical numerical techniques for solving ordinary differential equations can then be employed.
Examples of this method of solution applied to equations whose mathematical structure closely resembles the one presented in the present paper
can be found in~\cite{[FFG09],[G12],[GG14]}.
Notice should be made that the resulting system of ordinary differential equation can be usually very large for a network where in each node various
functional subsystems interact. Indeed this is the case of the developments on the modeling of the criminality dynamics reported in the following
section. Although parallel computing can be used to alleviate the computational burden~\cite{[FGG11a],[FGG11b]},
alternative numerical approaches are certainly of interest.

According to the authors' bias, Monte Carlo simulation~\cite{[DPR04],[PT13]} methods are ideally suited for solving the class of equations under
consideration.
The basic idea consists in representing the distribution function by a number of computational particles.
These particles move through the network, unless migration dynamics is disregarded, and interact according to stochastic rules derived from the
kinetic equation, Eqs.~\eqref{eq:structure}. The macroscopic fields are obtained through weighted averages of the particle properties.
Compared to deterministic methods of solution, the application of stochastic methods provides some
important advantages such as the computational efficiency and the possibilities to easily account for sophisticated individual decision processes.

It is worth noticing that a distinction is sometimes made between simulation and Monte Carlo, being the former the direct coding of a natural
stochastic process while the latter the solution by probabilistic methods of non probabilistic problems.
However, often such a distinction cannot be maintained. A relevant example is provided by the Direct Simulation Monte Carlo (DMSC), which is by far
the most popular simulation method for classical kinetic theory applications.
Although it has been introduced based on physical reasoning~\cite{[B94]},
DSMC has been later proved to converge, in a suitable limit, to the solution of the Boltzmann
equation~\cite{[W92]}. Therefore the same computer code can be regarded simultaneously as a direct physical transcription of the dynamics of a
rarefied gas and as a stochastic solution of the Boltzmann equation.


\subsection{On the use and abuse  of the so-called big data}

This subsection aims at showing how the systems approach can contribute to the use of the so-called big data toward predictive purposes on an ongoing observed
dynamics. As it is known, thanks to the advent of information and communication technology, as well as to the increasing amount of (big) data is collected everyday.  These data contain information about the behavior of individuals (such as their movements, their networks of friends, the sharing  of opinions via social networks, etc) and are increasingly becoming a valuable source of information to validate models around the behaviors of individuals. In general, the term \textit{big data} is used to define data sets so large or complex that traditional data processing applications are inadequate.
An important reference is given by the report~\cite{[CHEN13]} by  paper~\cite{[DSJ15]}, while additional information can be obtained by the reports posted in the WEB sites~\cite{[REPORT]}.

The problem consists in understanding how these data can be used for predictive purposes  or for model validation. Currently, the various methods to treat these large amounts of data are being developed define  an emerging - so called - data science which aims at improving the  decision process toward cost reductions and reduced risk. Rather than treating the technical problem of data compression, we suggest, as a possible perspective, the use of big data within the systems approach proposed in this paper focused on the analysis of an observed dynamics where the main problem consists in selecting specific actions to drive the dynamics toward an optimal trend. This target refers to a large number of data collected in a repository database. The approach can be developed along the following sequence of actions:

\begin{enumerate}

\item Classification of the data according to the specific features of the systems approach, namely by functional subsystems and activity variables. These features are used to compare different dynamics independently on external actions.

\vskip.1cm \item  Define an appropriate metrics suitable to compare the specific features of the system under consideration with similar dynamics stored in the aforementioned database. Such a metrics defines the distance between the selected dynamics $S_d$ and the stored one $S_s$: $||S_d - S_r||$, where, after appropriate normalization, $||S_d - S_r|| = 0$ means maximal similarity, while $||S_s - S_r|| = 1$ means maximal distance.

\vskip.1cm \item  Select in the big data repository the  closest $S_s$ to the system under consideration and provide for each of them a score $V \in [0,1]$, where $V = 1$ is the best score, while $V = 0$ means the worst one.

\vskip.1cm \item  Select in the database the optimal dynamics, namely the optimal external actions, by using  a convex combination of the distance and the score. For instance:
$$
J = \alpha \,(1 - ||S_d - S_r||) + (1 - \alpha)\, V,
$$
where the optimal selection corresponds to the highest value of $JV$.

\vskip.1cm \item Apply to the real systems the external actions corresponding to the dynamics in the database corresponding to the system selected according to item (4).

\end{enumerate}

\begin{Remark}
The milestones (1)--(5) allow to apply a decision process related to the selection of actions appropriate to control the dynamics of systems by selection actions appropriate to control an ongoing dynamics. The actual expression of of the objective functional can be further revised  by enriching it of additional variables according to the specific system which is treated. The method is general and can go beyond applications in social dynamics.
\end{Remark}


\section{Reasonings on a Case Study}
\label{criminality}

Let us now show how the general systems approach proposed in the preceding section can be applied in a specific case study. 
In detail, we consider the onset and development  of criminality in a society, where this dynamics is contrasted by security forces,
such as intelligence and police.

The choice of the case study  refers to a mathematical literature, which is constantly growing due also to the impact that this topic has on the wellbeing and security of citizens. Useful references are given by the following essays~\cite{[BCD14],[FLL02],[FEL10],[HHM10],[ORM05]}, which are brought to the attention of the reader as examples without aim of completeness.

Different mathematical approaches have been used corresponding to different representation scale and tools to model interactions.
Nonlinear dynamical systems have been developed in~\cite{[NHP11]}, while the use of compartmental models~\cite{[BCGS14]} allows modeling transitions across what we have called functional subsystems, and  predicting, at least in some specific cases, tipping points. The approach by partial differential equations, which include not only number of individuals, but also their localization has been proposed in various papers, for example~\cite{[BWW14]}, where accounting for the localization includes the modeling of pattern formation of criminality and its propagation in the territory.
The dynamics of shocks and acceleration waves are studied in~\cite{[ST14]}.

This section illustrates how the hallmarks of the system theory previously discussed can be applied to the model presented in~\cite{[BCKS15]}, which shows how the contrast actions are related on one side to the social state of citizens, such as wealth and culture, and, on the other hand, to the ability of security forces based also on their training. More in general, the interplay between wealth distribution and unethical behaviors is object of growing interest of researchers, who operate in the field of social sciences~\cite{[KPM12]}, while it is becoming a new field of investigation of applied mathematicians~\cite{[BHT13],[DL14]}.

The contents  are presented in three subsections.
Subsection 6.1 briefly summarizes the model proposed in~\cite{[BCKS15]}.
Subsection 6.2 shows how a model with discrete states can be designed in agreement with Remark 3.6.
Finally Subsection 6.3  proposes some possible developments.

\subsection{On a model of social criminality}

The mathematical model studied in~\cite{[BCKS15]} accounts for the following interaction dynamics: Susceptibility of citizens to become criminals,  susceptibility of criminals to reach back
the state of normal citizens, learning dynamics among criminals, motivation/efficacy of security forces  to catch criminals, learning dynamics
among security and interplay with wealth dynamics.

In more details, the following features of the model proposed in~\cite{[BCKS15]}, are briefly analyzed:

\vskip.2cm \no \textbf{Functional subsystems and activity:} 
The following subsystems are considered: Citizens, criminals and security forces, while the activity variables expressed by them are, respectively, 
the following: Wealth, criminal ability and detective ability.

\vskip.2cm \no \textbf{Mathematical structure:} The mathematical structure used in~\cite{[BCKS15]} is the following:
\begin{eqnarray*}\label{eq:structure_c}
&& \partial_t f_i (t,u) = J_i[{\bf f}](t,u)=
 \nonumber \\
 && \,   =  \sum_{h,k=1}^3 \int_{D_h} \int_{D_k} \eta_{hk}(u_*, u^*)
 \mathcal{B}_{hk}^i (u_* \to u| u_*,u^*) f_h(t, u_*) f_k(t, u^*) \, du_* \, du^* \nonumber \\
&& \qquad - f_i(t,u) \sum_{k=1}^3 \,\int_{D_k} \, \eta_{ik}(u, u^*)\, f_k(t,u^*)\,  du^* \nonumber \\
&& \qquad + \int_{D_i} \mu_{i}(u_*,\mathbb{E}_i) \mathcal{M}_{i}(u_*\rightarrow u | u_*, \mathbb{E}_i) f_i(t,u_*) du_*  \nonumber \\
&& \qquad - \mu_{i}(u,\mathbb{E}_i) f_i(t,u),
\end{eqnarray*}
where $f_i:[0,T] \times D_i \to \RR_+, \, i  = 1, 2, 3$, while $\eta_{hk}(u_*,u^*)$ and   $\mu_{i}(u_*,\mathbb{E}_i)$ are
the encounter rates between a candidate $h$-particle and a field $k$-particle and between a candidate $i$-particle and the mean activity, respectively.
Moreover,  $\cB_{hk}^i(u_* \to u|u_*,u^*)$ and $\mathcal{M}_{i}(u_*\to u | u_*, \mathbb{E}_h)$ are the probability density for the
state transition of individual due to interactions between particles and between particle and the mean activity, respectively.
Such a structure is a particular case of the general one reported in Eq.(3.5)--(3.9).
It is worth noticing that external actions are not accounted for and space dynamics is disregarded.
Therefore this model refers to the simple case of a closed system which comprises only one node.

\vskip.2cm \no \textbf{Modeling encounter rates:} Citizens with closer social states interact more frequently; 
unexperienced lawbreakers tend to expose themselves more frequently; experienced detectives are more likely to chase less experienced criminals.

\vskip.2cm \no \textbf{Modeling individual based interactions:} Citizens are susceptible to become criminals, motivated by their wealth state, thus they can mutate into a new functional subsystem.  Criminals interact among themselves resulting in a dynamics by which less experienced criminals  mimic the more experienced ones.  Criminals chased by detectives  are constrained to step back decreasing their activity value as the price to be paid for being caught. At the same time, detectives gain experience from a well-done job increasing their activity. 
Due to this action, criminals are induced to return to the state of normal citizens with probability which increases with decreasing values of their level of criminality and increasing values of skill of detectives.

\vskip.2cm \no \textbf{Interactions between individuals and the whole functional subsystem:} Both criminals and detectives interact with  the mean value through  the mean-micro state distance within their own functional subsystem, in both cases the interaction rate increases with the distance between the individual state and the mean value. The dynamics is such that only those who are less experienced than the mean tend to learn and move towards it. 
Likewise, detectives show a trend toward the mean value.

For additional technical details on the terms modeling interactions, namely interaction rates and transition probability densities,
we refer the reader to~\cite{[BCKS15]}, where a number of simulations are carried out which indicate that an equal distribution of wealth leads to
a slow growth, or even a decrease of criminals, while the opposite trend is observed in the case of  unequal distributions.
Concerning the role of detectives the model shows that the expertise which they acquire due to training more than their number has a positive effect on the control of the number of criminals and their aggressive behaviors.

\subsection{On a model with discrete states}

As pointed out in the Remark~3.6, the use of models with discrete states is motivated by the difficulty of dealing with a continuous activity variable.
Moreover, in several cases, the discrete approach can simplify the modeling of interactions at the microscopic scale and even enriches the predictive
ability of the model.
This subsection aims at showing how a model with discrete states can be derived under the same
phenomenological assumptions proposed in the preceding subsection.

As in Section~6.1, the functional subsystems and related activity variables are the following:
$i=1:$  citizens and wealth;  $i=2:$  criminals and criminal ability;	$i=3:$  detectives  and their skill.
However, we here introduce a discrete activity variable, labeled by the subscript $\si=1, \ldots, m$ with $u_1=0$ and $u_m=1$. 
The abbreviation $i;\si$-particle is used for a particle belonging to the $i$-th functional subsystem with activity $u_\si$. The representation of the overall system is delivered by the discrete probability 
$f_i^\si = f_{i}^\si(t)\,: \, [0,T] \, \to \RR_+$, while weighted sums replace the weighted integrals to obtain variables at the macroscopic scale.

The actual derivation needs defining the interaction terms in the new context as well as a new mathematical structure. In detail:

\begin{itemize}
\item $\eta_{hk}^{pq}$ is  the encounter rate  between a candidate $h;p$-particle  and a field $k;q$-particle.

\vskip.2cm \item  $\mu_{ih}^p$ is the encounter rate between a candidate $h;p$-particle and the mean activity within its functional subsystem.

\vskip.2cm \item $\cB_{hk}^{pq}(i;\si)$ is the discrete  transition probability that a  candidate $h;p$-particle ends up into the state of the test $i;\si$-particle due the interaction with a field $k;q$-particle, such that

\vskip.2cm \item   $\mathcal{M}_{h}^p(i;\si)(u_p, \mathbb{E}_h)$ is the  transition probability that a candidate  $h;p$-particle  ends up into the state $i;\si$-particle due the interaction with the mean activity value $\mathbb{E}_h$. such that that
\end{itemize}

The transition probabilities satisfy, for all type of inputs, the following normalization
\begin{equation}\label{density}
\sum_{i=1}^3 \sum_{\si=1}^m \mathcal{B}_{hk}^{pq}(i;\si) = \sum_{p=1}^m \mathcal{M}_{h}^p(\mathbb{E}_h) = 1.
\end{equation}

\vskip.2cm \no {\bf Derivation of a mathematical structure}
The balance of particles in the elementary volume of the space of micro-states leads to the following structure:
\begin{eqnarray}\label{eq:structure_c_bis}
&& \partial_t f_{i}^{\si} (t) = J_{i}^{\si}[{\bf f}](t)=
 \nonumber \\
 && \,   =  \sum_{h,k=1}^3  \sum_{p,q=1}^m  \eta_{hk}^{pq}
 \mathcal{B}_{hk}^{pq}(i;\si) f_h^p(t) f_k^q(t, u^*) - f_{i}^{\si}(t) \sum_{k=1}^3 \sum_{q=1}^m  \, \eta_{ik}^{pq}\, f_k^q(t,u^*)\nonumber \\
&& \quad +  \sum_{h=1}^3  \sum_{q=1}^m   \mu_{ih}^q \mathcal{M}_{i} f_i^q(t) - \sum_{q=1}^m  \mu_{i;\si}^q(u,\mathbb{E}_i) f_{i}^\si(t).
\end{eqnarray}

\vskip.2cm
The derivation of the mathematical model is obtained within the general structure given by Eq.(\ref{eq:structure_c})
by particularizing the terms $\eta$, $\mu$, $\mathcal{B}$ and $\mathcal{M}$.

\vskip.2cm \no {\bf Encounter rate}
Let us consider  the encounter rates $\eta_{hk}^{pq}$ and $\mu_h^q$. The modeling approach is based on heuristic assumptions that let us quantify the frequency
of interactions, depending on the micro-states and distribution functions of the interacting particles. The result, where candidate particles are represented by a square, while field particles are represented by a circle, is the following:

\begin{itemize}

\item \framebox{\footnotesize 1}  $\leftrightarrow$ \textcircled{\footnotesize 1}: Closer social states interact  more frequently $\eta_{11}^{pq}=\eta^{0} \left(1-|u_p-u_q|\right)$;

\vskip.2cm \item \framebox{\footnotesize 2}  $\leftrightarrow$ \textcircled{\footnotesize 2}:   Experienced lawbreakers  are more expected to  expose themselves
$\eta_{22}^{pq} = \eta^{0} (u_p+u_q)$;

\vskip.2cm  \item \framebox{\footnotesize 2}  $\leftrightarrow$ \textcircled{\footnotesize 3}:  Experienced detectives  are more likely to less experienced criminals {\em hunt}  $\eta_{23}^{pq} = \eta^{0} \big((1-u_p) + u_q\big)$;

\vskip.2cm  \item \framebox{\footnotesize 2}  $\leftrightarrow$ $\mathbb{E}_2$: Criminals  interact with the mean value through the mean-micro state distance $\mu_{2}(u_*,\mathbb{E}_2)= \mu^{0}|u_* - \mathbb{E}_2|$;

\vskip.2cm  \item \framebox{\footnotesize 3}  $\leftrightarrow$ $\mathbb{E}_3$: Detectives  interact with the mean value through the mean-micro state distance $\mu_{3}(u_*,\mathbb{E}_3) = \mu^{0}|u_*-\mathbb{E}_3|$.

\end{itemize}

\vskip.2cm \no {\bf Transition probability:}  The social structure of the population $i=1$ is assumed to be fixed,  namely, the time interval is sufficiently short, that the wealth distribution is  constant in time.

\begin{itemize}

\item  The probability that  citizens  become a criminals, increases with  their decreasing wealth state referred to that of other citizens:
 $$
    \cB_{11}^{pq}(21) = \alpha_A \, (1-u_p)u_q, \quad  \cB_{11}^{pq}(11) = ( 1 - \alpha_A \, (1-u_p)u_q).
    $$

\vskip.2cm \item Less experienced criminals  mimic the more experienced ones, moreover
also interaction with less experienced lawbreakers increases the level of criminality:
$$
\cB_{22}^{pq}(2(p+1)) = \beta(1-u_p) u_q, \quad  \cB_{22}^{pq}(2p) = (1 - \beta(1-u_p)u_q).
$$

\vskip.2cm \item Criminals are constrained by detectives  to step back decreasing their activity value
as the price to be paid for being caught:
$$
p = 1: \, \cB_{23}^{pq}(1,1) = \alpha_B (m - 1) u_p, \quad  \cB_{23}^{pq}(2,1) = (1 - \alpha_B (m - 1) u_p);
$$
$$
p > 1: \, \cB_{23}^{pq}(2,p - 1) = \alpha_B  (m - p) u_p, \quad  \cB_{23}^{pq}(2,p) = (1 - \alpha_B (m - 1) u_p).
$$

\vskip.2cm \item Detectives gain experience from their activity and increase their skill:
$$
\cB_{32}^{pq}(3,p+1) = \gamma u_q(1-u_p), \quad (1 - \gamma u_q(1-u_p)).
$$
\vskip.2cm \item  The level criminality and the skill of detectives show a trend toward their respective  mean value:
$$
\mathcal{M}_{i} > p: \mathcal{M}_{i}(p + 1)  = \lambda(\mathbb{E}_i - p)(m - p), \quad \mathcal{M}_{i}(p)  = 1 -  \lambda(\mathbb{E}_i - p)(m - p);
$$
$$
\mathcal{M}_{i} < p: \mathcal{M}_{i}(p - 1)  =  \lambda (p - \mathbb{E}_2)(m - p), \quad \mathcal{M}_{u}(2,p)  = 1 -  \lambda(p - \mathbb{E}_i)(m - p),
$$
where $i,2$.
\end{itemize}

Therefore the model needs five phenomenological parameters, namely  $\alpha_A$ is the susceptibility of citizens to become criminals;
  $\alpha_B$ is the susceptibility of criminals to reach back the state of normal citizens; $\beta$ refers to the learning dynamics among criminals;
 $\gamma$  models the motivation/efficacy of security forces  to catch criminals; and
$\lambda$  the learning (attraction) dynamics with respect to the mean value.

Let us now briefly comment the result of simulations focusing on the role of the \textit{wealth distribution}. The approach allows to investigate not only the role of macroscopic quantities such as the mean wealth, but also its distribution at equal mean value. More in detail, referring to the influence of the \textit{mean wealth} the following result is obtained:

\begin{itemize}

\item Decreasing mean wealth of the society $\Longrightarrow$ Increasing the  number of criminals and increasing their criminal ability;

\vskip.2cm \item Increasing mean wealth of the society $\Longrightarrow$ Decreasing the  number of criminals and decreasing their criminal ability;

\end{itemize}

Let us now focus on the influence of the shape \textit{shape of wealth distribution} the following result is obtained:

\begin{itemize}

\item   Equal distribution in a poor society   $\Longrightarrow$   Slow growth of the number of criminals;

\vskip.2cm \item  Equal distribution  in a wealthy society $\Longrightarrow$  Fast decrease of the number of criminals; 

\vskip.2cm \item  Unequal distribution in a poor society   $\Longrightarrow$  Fast growth of the number of criminals;

\vskip.2cm \item    Unequal distribution in a wealthy  society  $\Longrightarrow$ Slow decrease of the number of criminals;

\end{itemize}

\subsection{Possible developments}

As already mentioned, the model has been derived under quite limited assumptions on the number of functional subsystems and in absence of a dynamics over
the network. Both aspects of the dynamics are important as a higher number of levels of criminal behaviors  and detective expertise are definitely closer to reality, while the effect of networks can have an important influence on the overall dynamics of the system. Indeed, social aspects of the interaction between
wealth dynamics and criminal  behaviors are object of intense studies by experts in the field of social science~\cite{[FLL02],[HHM10],[HP93]}.

 We can observe that a number of recent papers, as examples~\cite{[BWW14],[SDP08]}, study the concentration, and related properties, of hot-spots of criminals. A more detailed information can be achieved by subdividing the territory into a network of interconnected sub-territories and subsequently in studying the migration and concentration of criminals in certain zones. A challenging objective consists in deriving macroscopic scale equations from the underlying description at the microscopic scale. More in details,  hyperbolic models with the feature of finite speed of propagation can be obtained~\cite{[BBNS07]},
 as well as degenerate parabolic models~\cite{[BBNS12]}. Previous applications to vehicular traffic~\cite{[BBNS14]} can be developed also in the case of social dynamics.

 However, according to the authors' bias the study of the dynamics over networks is definitely more important also in view of its practical
 applications.

 As an example, if a certain territory, as an example a city, is subdivided in a number of subareas and if migration phenomena across them are
 modeled, the concentrations of criminality can be precisely identified and thus the planning of police forces distribution can be properly organized.


\section{Some Reasonings on the Validation of Models}
\label{validation}

Focusing on a case study, it has been shown how the proposed approach can be applied to derive specific models.
However, the general problem of model validation has not yet found an exhaustive reply not only  in the existing literature, but also in this present
paper. Therefore, it is necessary tackling  the last key question:

\vskip.1cm
\begin{center}
  {\dc How and how far mathematical models can be validated?}
\end{center}
\vskip.1cm

Although not yet exhaustive, a partial  answer will be given by  going back to the key problems posed in \cite{[BL12]}.
In general, validation  requires that models can reproduce empirical data at the scale used to observe and represent the phenomena they describe. Moreover they should reproduce, at least at a qualitative level, emerging collective behaviors.

The reasonings presented in the following do not claim to provide, an exhaustive reply on this challenging topic, but simply a focus on some key problems looking ahead to research perspectives.

\vskip.1cm \no  \textbf{i) Multiscale problems:} Models might be designed at a scale different from that used for the observation and detection of empirical data. Generally, models refer to the microscopic scale, while only collective behaviors at the macroscopic scale  are observed.

\vskip.1cm \no \textbf{ii) Dynamics of complex systems:} Complex systems are characterized by collective behaviors which are subject to large deviations. However, qualitative shapes are in several cases preserved. Models are supposed to reproduce all observed emerging behaviors and, hopefully, also to predict, in some cases, behaviors that were previously unnoticed.

\vskip.1cm \no  \textbf{iii) From interactions to collective dynamics:} Collective dynamics are determined by interactions at the microscopic scale, which in turn are ruled by the strategy that interacting entities are able to express. Specific experiments should be designed to understand the dynamics of individual social interactions.

\vskip.1cm \no  \textbf{iv) Evolution of rules of interactions with  onset of rare events:}  A mathematical approach to modeling should consider that individual behavioral rules and strategies are not, in most cases, constant in time due to the evolutionary characteristics of living complex systems. For instance, interactions among the individuals can change in time depending on several factors, including the global state of the system. Such variability can be a source of unpredictable events. Networks can enhance rapid spreads across nodes  as a sort of \emph{domino effect}.

\vskip.1cm

\begin{Remark}
Taleb's definition~\cite{[TAL07]} of the so-called ``Black Swan'' was already presented in the first section. It is worth stressing that this author is a rare example of
researcher moving against the main stream of the traditional approaches, generally focused on well-predictable events. As such, it had an important impact on the quest for new research perspectives. In line with the approach followed in~\cite{[BHT13]}, some speculations can be developed to provide an answer to the following  key topic which deserves further speculations,
that is the interplay among different dynamics an important ingredient for the onset of a non predictable event.
\end{Remark}

All reasonings presented in this subsection do not directly contribute to the validation of models,  but they  enlighten the conceptual difficulties to be tackled. Indeed, the present state of the art does not yet provide well defined validation rules. Therefore, additional speculations can be presented by analyzing how far the theory proposed in this paper can provide a constructive answer to the key problems
proposed in the book \cite{[BL12]} and reported in Section 2. The reasonings proposed in the following correspond to the personal opinion of the authors of this present paper.

\begin{enumerate}

 \item \textbf{Separation between homogeneous  and heterogeneous behaviors}: The separation between these two types of dynamics, which correspond to homophyly and heterophily,  is modeled by a threshold, which separates consensus from dissent. According to the approach of this paper, the threshold depends on the distance between the interacting entities. In some cases, for example opinion formation, small distances induce consensus, while the opposite trend is observed in the case of social competition.  It is worth mentioning that such a threshold cannot be considered a constant quantity, as it depends not only on external actions, but also to the overall state of the system.

\vskip.1cm \item  \textbf{Dynamics toward consensus}: Understanding the specific features that promote consensus has been a constant subject of research activity by mathematicians and physicists, examples are given by  \cite{[HY09],[MT14]}. A recent survey \cite{[MT14]} provides a deep analysis on this topic. The approach presented in this paper links the dynamics toward consensus (cooperation) to the threshold already mentioned in the preceding item. Therefore, the trend increases (or decreases) according to the dynamics and external actions, which modifies such a threshold. Papers \cite{[ABT13],[DL14]} present specific applications related to this topic.

\vskip.1cm \item  \textbf{Dynamics over networks:} The main observation concerning the modeling of networks is related to the way by which the partition into functional subsystems has been organized. As already mentioned, this process is not developed according to universal rules, but it is related on the specific investigation which is pursued. As a consequence, the organization into functional subsystems and hence into a network  is not fixed for all times as the dynamics of subsystems might induce changes in the network to be viewed as a dynamical system \cite{[KSAX10]}, where new subsystems can appear, while others disappear.

\vskip.1cm \item  \textbf{Measure of centrality of networks}: This topic remains open after the contribution of this present paper, which is limited to analyze how the network as a whole has an influence over the dynamics of each functional subsystem. A quantitative indication has been related to the influence of the mean value. However, this topic deserves additional studies to understand the role of higher order moments as well as the higher or lower influence of one node with respect to the others. This aspect can be accounted for by the interaction rate, while an interesting objective of investigation consists in understanding how far the action of the network can modify the dynamics within each node.

\vskip.1cm \item  \textbf{Unpredictability and chaotic dynamics}: This topic deserves attention and should be developed according to the class of equations used in the modeling approach. Stability and chaotic dynamics are rapidly related to models stated in terms of ordinary differential equations. Focusing on the stochastic equations proposed in this paper, a natural investigation is the analysis, by appropriate entropy functions \cite{[BALE90]}, of the asymptotic trend by understanding whether shows a trend toward a deterministic behavior or toward an increasing variance. At present only computational results support this study \cite{[BCKS15],[BHT13]}, while analytic studies definitely deserve future contributions.

\end{enumerate}


\section{Looking ahead to Research Perspectives}
\label{perspectives}

This paper does not simply focus on a survey of the existing literature but it proposes a systems approach to social sciences. Such approach  is  developed in two steps. The first step consists in the derivation of a general mathematical structure suitable to capture the main features of behavioral social systems and the second step in deriving specific models by specializing the parameters of the structure according to each specific system under consideration. This approach includes the dynamics over networks and migration dynamics
across the various nodes of the network.

Looking ahead to research perspectives, firstly we point out the advantages of the aforementioned systems approach and then we critically analyze the
shades to be hopefully enlightened by future research activity.

According to the authors' bias, the main advantage  is the broad generality of the mathematical structure, which appears to be consistent with the
main features of living systems. Moreover, it has been verified that the key features proposed in  the book \cite{[BL12]} have been effectively treated and a constructive answer has been given. Therefore, it can be stated that such a structure offers a tool useful for the
applications as it has been shown in the case study reviewed in Subsection 5.2.

Possible improvements of the approach can be obtained by studying into more depth five specific topics,
selected according to the authors' bias and are treated in the following.
These topic share a common focus on the modeling interactions according to the idea that some preliminary reasonings on this topic can hopefully contribute to future research activity.

\vskip.2cm \no \textbf{8.1 Modeling interactions toward a socio-mathematical theory:} A deep understanding of the dynamics of interactions can provide an important contribution to the study of collective dynamics.
The mathematical structure proposed in this paper can transfer, as we have seen, the individual behavior to the collective dynamics. Theoretical contributions from social sciences can  provide a theoretical approach to such a modeling. This challenging objective can establish a constructive link between mathematics and social sciences by linking the mathematical structure to theories of social systems with some analogy to the interaction between physics and mathematics which, in the past century, generated the so-called mathematical physics. The final objective consists in the development of a social-mathematical theory suitable to offer a vision broader than that offered by single models.

\vskip.2cm \no \textbf{8.2 Learning dynamics:} This concept has been mentioned several times in the paper. Indeed, it plays a key role on the modeling of interactions, as active particles modify their rules of interactions according to what they learn from other particles and has been already object of attentions in various applications as we have seen in Subsection 5.2. Accordingly, the theoretical approach to learning is not simply collecting and organizing a large set of information, as it needs modeling the actual dynamics by which individuals ``learn'' from the others. This approach has been recently developed in \cite{[BDG15]} based on the idea that understanding the dynamics can contribute to modeling interactions. Particularly important is the modeling of learning dynamics across networks \cite{[AO11]}, where individuals perceive the overall state of the functional subsystems and learn out of this knowledge with a direct influence over the social dynamics.

\vskip.2cm \no \textbf{8.3 Multiple dynamics:} Some recent papers, see among others,  \cite{[BHT13],[DL14]}, have shown that the interactions of different dynamics can explain social phenomena that cannot be foreseen when each dynamics is taken into account separately from the other. Interesting examples are treated \cite{[BHT13],[DL14]}. This literature motivates further research in this field, where a  challenging objective consists in understanding how the interaction rules of one dynamics can be extended to  the case of multiple dynamics. Generally, a factorization of the transition probability density is assumed to simplify mathematical models, but this heuristic assumption  does not find any theoretical support.

\vskip.2cm \no \textbf{8.4 Additional developments on dynamics over networks:} This paper has already proposed some perspective ideas on the modeling of networks by selecting various types on interactions either within the same node or across them. Two scales have been used, namely the microscopic one related to interactions involving active particles
and micro-macro interactions between active particles and nodes as a whole. However, some limits still need to be properly understood.
Examples include the limitation to first order moments in the micro-macro interactions and the assumption, in the micro-interactions, that each particle interact with all particles in a node.
The use of higher order moments can contribute to model more precisely interactions. As an example the interaction rate could be modeled based on the conjecture that FSs featured by a high second order moment (somehow corresponding to energy) have a stronger influence on the FSs with low second order moment, namely the encounter rate grows with the difference between the relative moments. Moreover, interactions might be selective and limited to a low number of particles due to topological, rather than metric, interactions. This concept is proposed in \cite{[BKS13]}.

\vskip.2cm \no \textbf{8.5 Analytic problems:} The application of models to the study of real problems generates some challenging analytic problems,
most of them have already been mentioned in the preceding sections focused on the initial value problem for specific models.
We have already stressed that the proof of solutions to such a problem simply needs the application of standard techniques.
Paper \cite{[ADF12]} has shown under which assumptions, reasonable from the view point of applications, existence and uniqueness can be proved by
almost straightforward applications of fixed point theorems \cite{[ADF12]}. In fact, the structure of the equations guarantees local existence,
under appropriate Lipschitz continuity assumptions of the integral operators.
Moreover,  the solutions can be prolonged for arbitrary long times as the aforementioned structure assure preservation of the $L_1$ norm.

On the other hand, the study of the asymptotic trend of the solutions is still waiting for a satisfactory answer.
Numerical evidence suggests that for each initial condition,
solutions tend, asymptotically in time, to a configurations which appears to be unique and stable. However, this phenomenological observation is not followed by a proof. A simple reasoning observes that this specific class of systems need a new definition of equilibrium. Indeed asymptotic configurations depend on the shape of the initial condition. Moreover, we stress that the study of systems subject to external actions is still looking for detailed analysis related to existence and control of solutions.

\vskip.2cm Of course the five topics, presented above, do not claim to offer an exhaustive panorama, but simply promote new research  initiatives as well as contributions of interested readers.
It is not surprising  that all of them focus on interactions. In fact, all different mathematical approaches to active particles, such as methods of Brownian particles \cite{[RBE12]} or kinetic theory methods \cite{[PT13]}, have to tackle the conceptual difficulties of modeling interactions among individuals. Which is the first research topic presented in this section.


\vskip.5cm  \noindent \textbf{Acknowledgments:} The research leading to these results has received funding from the European Union's Seventh Framework Programme (FP7/2007--2013) under grant agreement number (SAFECITI). Project full title: ``Simulation Platform for the Analysis of Crowd Turmoil in Urban
Environments with Training and Predictive Capabilities''. This publication reflects the views only of the authors, and the Commission cannot be held responsible for any use which may be made of the information contained therein.



\begin{thebibliography}{10}

\bibitem{[ABO10]}
\newblock D.~Acemoglu, K.~Bimpikis, and A.~Ozdaglar,
\newblock Dynamics of information exchange in endogenous social networks,
\newblock {\em National Bureau of Economic Research}, n.16410, (2010).

\bibitem{[AO11]}
\newblock D. Acemoglu and A. Ozdaglar,
\newblock Opinion dynamics and learning in social networks,
\newblock {\em Dyn. Games Appl.}, \textbf{1}  3-49, (2011).

\bibitem{[AR06]}
\newblock D.~Acemoglu  and J.A.~Robinson,
\newblock  \textbf{Economic Origins of Dictatorship and Democracy},
\newblock Cambridge University Press, Cambridge, (2006)

\bibitem{[AR06B]}
\newblock D.~Acemoglu  and J.A.~Robinson,
\newblock  Economic backwardness in political perspective,
\newblock {\em American Political Science Review}, \textbf{100(1)} 115-131, (2006).

\bibitem{[ATV11]}
\newblock D.~Acemoglu, D.~Ticchi and A.~Vindigni,
\newblock Emergence and persistence of inefficient states,
\newblock  {\em J. Eur. Econ. Assoc}, \textbf{9} 177-208, (2011).

\bibitem{[ABH13]}
\newblock S.~Ahn, H-O. Bae, S-Y.~Ha, Y.~Kim, and H.~Lim,
\newblock Application of flocking mechanicsm to the modeling of stochastic volatility,
\newblock  {\em Math. Models Methods Appl. Sci.},  \textbf{23} 1603-1628, (2013).

\bibitem{[AM09]}
\newblock G.~Ajmone Marsan,
\newblock  New paradigms towards the modelling of complex systems in {B}ehavioral {E}conomics,
\newblock {\em Math. Comp. Modelling}, \textbf{50} 584-597, (2009).

\bibitem{[ABE08]}
\newblock G.~Ajmone~Marsan, N.~Bellomo, and M.~Egidi,
\newblock Towards a mathematical theory of complex socio-economical systems by functional subsystems representation,
\newblock {\em Kinet. Relat. Models},  \textbf{1(2)} 249-278, (2008).

\bibitem{[ABT13]}
\newblock G.~Ajmone~Marsan, N.~Bellomo, and A.~Tosin,
\newblock   \textbf{Complex Systems and Society -- Modeling and Simulations},
\newblock {\em Springer Briefs}, Springer, New York, (2013).

\bibitem{[ABHT14]}
\newblock G.~Ajmone~Marsan, N.~Bellomo, M.A. Herrero, and A. Tosin,
\newblock From  five key questions to a  system sociology theory - Predicting the unpredictable : hunting black swans,
\newblock  \textbf{Tipping Points : Modelling Social Problems \& Health},  By J.J.~Bissell, C.C.S.~Caido, M.~Goldstein and B.~Straughan Eds., Chapter 7, Wiley, London, (2015).

\bibitem{[AB02]}
\newblock R.~Albert and A.L.~Barab\'asi,
\newblock Statistical mechanics of complex networks
\newblock {\em Rev. Modern Phys.},  \textbf{74} 47-97, (2002).

\bibitem{[ANT07]}
\newblock G.~Aletti, G.~Naldi, and G.~Toscani,
\newblock First-order continuous models of opinion formation
\newblock {\em SIAM Journal on Applied Mathematics}, \textbf{67} {837-853}, (2007).

\bibitem{[A01]}
\newblock V.V~Aristov,
\newblock \textbf{Direct Methods for Solving the Boltzmann Equation and Study of Nonequilibrium Flows},
\newblock Springer-Verlag, New York, (2001).

\bibitem{[ABL00]}
\newblock L.~Arlotti, N.~Bellomo, and  M.~Lachowicz,
\newblock Kinetic equations modelling population dynamics
\newblock {\em Transp. Theor. Statist. Phys.},  \textbf{29} 125-139, (2000).

\bibitem{[ADF12]}
\newblock L.~Arlotti, E.~De~Angelis, L.~Fermo, M.~Lachowicz, and N.~Bellomo,
\newblock On a class of integro-differential equations modeling complex systems with nonlinear interactions,
\newblock {\em Appl. Math. Lett.},  \textbf{25} 490-495, (2012).

\bibitem{[ADL97]}
\newblock W.B.~Arthur, S.N.~Durlauf, and D.A.~Lane, Editors,
\newblock  \textbf{The Economy as an Evolving Complex System II}, volume XXVII of
\newblock {\em Studies in the Sciences of Complexity}, Addison-Wesley, (1997).

\bibitem{[BO12]}
\newblock F.~Bagarello and F.~Oliveri
\newblock A phenomenological operator description of interactions between populations with applications to migration,
\newblock {\em Math. Models Methods Appl. Sci.},  \textbf{22} 471-492, (2012).

\bibitem{[BALE90]}
\newblock K.D.~Baley,
\newblock  \textbf{Methods of Social Research and Entropy Search},
\newblock Suny Press, (1990).

\bibitem{[BALE94]}
\newblock K.D.~Baley,
\newblock  \textbf{Sociology and New Systems Theory - Toward a Theoretical Syntesis},
\newblock Suny Press, (1994).

\bibitem{[BALL12]}
\newblock P.~Ball, Ed.,
\newblock  \textbf{Why Society is a Complex Matter},
\newblock Springer-Verlag, (2012).

\bibitem{[BAR12]}
\newblock A.L.~Barab\'asi,
\newblock  \textbf{The Science of Networks}
\newblock Perseus, Cambridge MA, (2012).

\bibitem{[BBV06]}
\newblock A.~Barrat, M.~Bath\'elemy, and A.~Vespignani,
\newblock \textbf{The Structure and Dynamics of Networks},
\newblock Princeton University Press, Princeton, (2006).

\bibitem{[BAT13]}
\newblock M.~Batty,
\newblock \textbf{The New Science of Cities}
\newblock The MIT Press, Boston, (2013).

\bibitem{[BT05]}
\newblock M.~Batty and P.M.~Torrneds
\newblock  Modelling and prediction in a complex world,
\newblock  {\em  Futures}, \textbf{37} 745-766, (2005).

\bibitem{[BFP09]}
\newblock U.~Bastolla, M.~A. Fortuna, A.~Pascual-Garc\'ia, A.~Ferrera, B.~Luque, and J.~Bascompte,
\newblock The architecture of mutualistic networks minimizes competition and increases biodiversity.
\newblock {\em Nature},  \textbf{458} 1018-1020, (2009).

\bibitem{[BEIN06]}
\newblock E.~Beinhocker,
\newblock  \textbf{The Origin of Wealth: Evolution, Complexity and the Radical Remaking of Economics},
\newblock Random House, (2006).

\bibitem{[BEIN12]}
\newblock E.~Beinhocker,
\newblock  \textbf{Complex new world: Translating new economic thinking into public policy},
\newblock The Institute for New Economic Thinking at the Oxford Martin School, (2012).

\bibitem{[BB15]}
\newblock N.~Bellomo and A.~Bellouquid,
\newblock  On multiscale models of pedestrian crowds from mesoscopic to macroscopic,
\newblock {\em Comm. Math. Sciences},  \textbf{13(7)}  1649--1664, (2015).

\bibitem{[BBNS07]}
\newblock  N. Bellomo, A. Bellouquid, J. Nieto, and J. Soler,
\newblock {Multicellular biological growing systems: Hyperbolic limits towards macroscopic description},
\newblock {\em Math. Models Methods Appl. Sci.}, \textbf{17} 1675-1692, (2007).

\bibitem{[BBNS12]}
\newblock  N. Bellomo, A. Bellouquid, J. Nieto, and J. Soler,
\newblock On the Asymptotic Theory from Microscopic to  Macroscopic Tissue Models: an Overview with Perspectives,
\newblock   \textit{Math. Models Methods Appl. Sci.},  \textbf{22}, paper n.~1130001, (2012).

\bibitem{[BBNS14]}
\newblock  N.~Bellomo, A. Bellouquid, J. Nieto, and J. Soler,
\newblock  On the multi scale modeling of vehicular traffic: from kinetic to hydrodynamics,
\newblock \textit{Discr. Cont. Dyn. Syst. Series B}, \textbf{19} 1869-1888, (2014).

\bibitem{[BCKS15]}
\newblock N.~Bellomo, F.~Colasuonno,  D.~Knopoff, and J.~Soler,
\newblock From a Systems Theory of Sociology to Modeling the Onset and Evolution of Criminality,
\newblock {\em Netw. Het. Media},  \textbf{10(3)}  421-441, (2015).

\bibitem{[BC13]}
\newblock N.~Bellomo and V.~Coscia,
\newblock Sources of nonlinearity in the kinetic theory of active particles with focus on the formation of political opinions,
\newblock {\em Contemporary Mathematics Series of the American Mathematical Society}, E.~Mitidieri, V.D.~Radulescu, and J.~Serrin, Eds., American Mathematical Society, Philadelphia, \textbf{594} 99-114, (2013).

\bibitem{[BE15]}
\newblock N.~Bellomo, A.~Elaiw,  A.M.~Althiabi, and A.~Alghamdi,
\newblock On the interplay between mathematics and  biology  hallmarks  toward a new systems biology,
\newblock  \emph{Phys. Life Rev.}, \textbf{12} 44-64, (2015).

\bibitem{[BB15b]}
\newblock N.~Bellomo and L.~Gibelli,
\newblock Toward a mathematical theory of behavioural-social dynamics for pedestrian crowds,
\newblock   \textit{Math. Models Methods Appl. Sci.},  \textbf{25} (2015), to appear.

\bibitem{[BHT13]}
\newblock N.~Bellomo, M.A.~Herrero, and A.~Tosin.
\newblock On the dynamics of social conflicts looking for the Black Swan,
\newblock {\em Kinet. Relat. Models},  \textbf{6(3)} 459-479, (2013).

\bibitem{[BKS13]}
\newblock N.~Bellomo, D.~Knopoff,  and J.~Soler,
\newblock On the difficult interplay between life ``complexity'' and mathematical sciences,
\newblock {\em Math. Models Methods Appl. Sci.},  \textbf{23} 1861-1913, (2013).

\bibitem{[BS12]}
\newblock N.~Bellomo and J.~Soler,
\newblock On the mathematical theory of the dynamics of swarms viewed as a complex system,
\newblock {\em Math. Models Methods Appl. Sci.},  \textbf{22} paper n.~1140006, (2012).

\bibitem{[BDF12]}
\newblock A.~Bellouquid, E.~De Angelis, and L.~Fermo,
\newblock Towards the modeling of vehicular traffic as a complex system: A kinetic theory approach,
\newblock {\em Math.  Models Methods Appl. Sci.},   \textbf{22}, paper n~.1140003 (2012).

\bibitem{[BDK13]}
\newblock A.~Bellouquid, E.~De Angelis and D.~Knopoff,
\newblock From the modeling of the immune hallmarks of cancer to a black swan in biology,
\newblock {\em Math. Models Methods Appl. Sci.},  \textbf{23} 949-978, (2013).

\bibitem{[BDEL06]}
\newblock A.~Bellouquid, and M.~Delitala,
\newblock \textbf{Modelling Complex Biological Systems - A Kinetic Theory Approach},
\newblock Series: Modeling and Simulation in Science, Engineering and Technology, Birkh\"auser, Boston, 2006.

\bibitem{[BCD14]}
\newblock  B.~Berenji, T.~Chou, and M.~D'Orsogna,
\newblock \emph{Recidivism and rehabilitation of criminal offenders: A carrot and stick evolutionary games}
\newblock PLOS ONE,  \textbf{9} paper n.~885531, (2014).

\bibitem{[BWW14]}
\newblock H.~Berestycki, J.~Wei, and M.~Winter
\newblock Existence of symmetric and asymmetric spikes of a crime hotspot model,
\newblock {\em SIAM J. Math. Anal.}, \textbf{46} 691-719, (2014).

\bibitem{[BD04]}
\newblock M.L. Bertotti and M.~Delitala,
\newblock From discrete kinetic and stochastic game theory to modelling complex systems in applied sciences,
\newblock {\em Math. Models Methods Appl. Sci.},  \textbf{14} 1061-1084, (2004).

\bibitem{[BD08]}
\newblock M.L. Bertotti and M.~Delitala
\newblock On the existence of limit cycles in opinion formation processes under  time periodic influence of persuadersm
\newblock {\em Math.  Models Methods Appl. Sci.}, \textbf{18(6)} 913-934, (2008).

\bibitem{[BM11]}
\newblock M.L.~Bertotti and G.~Modanese,
\newblock From microscopic taxation and redistribution models to macroscopic income distributions,
\newblock {\em Physica A}, \textbf{390} 3782-3793, (2011).

\bibitem{[BPS10]}
\newblock T.~Besley, T.~Persson, and D.M.~Sturm,
\newblock Political competition, policy and growth: Theory and evidence from the US,
\newblock Review of Economic Studies, \textbf{77} 1329-1352, (2010).

\bibitem{[BLHKW07]}
\newblock L.M.A.~Bettencourt, J.~Lobo, D.~Helbing, C.~Kohnert, and G.B.~West.
\newblock Growth, innovation, scaling, and the pace of life in cities,
\newblock {\em Proc. Natl. Acad. Sci. USA},  \textbf{104} 7301-7306, (2007).

\bibitem{[B94]} G.A.~Bird,
\newblock \textbf{Molecular Gas Dynamics and the Direct Simulation of Gas Flows},
\newblock Oxford University Press, (1994).

\bibitem{[BST09]}
\newblock M.~Bisi, G.~Spiga, and G.~Toscani,
\newblock Kinetic models of conservative economies with wealth redistribution,
\newblock {\em Comm. Math. Sciences}, \textbf{7} 901-916, (2009).

\bibitem{[BCGS14]}
\newblock J.J.~Bissell, C.C.S.~Caiado, M.~Goldstein and B.~Straughan,
\newblock Compartmental modelling of social dynamics with generalized peer incidence,
\newblock \emph{Math. Models Methods Appl. Sci.}, \textbf{24}  719-750, (2014).

\bibitem{[BDT99]}
\newblock E.~Bonabeau, M.~Dorigo, and G.~Theraulaz,
\newblock \textbf{Swarm Intelligence: From Natural to Artificial Systems},
\newblock Oxford University Press, Oxford, (1999).

\bibitem{[BL12]}
\newblock P.~Bonacich and P.~Lu,
\newblock \textbf{Introduction to Mathematical Sociology}
\newblock Princeton University Press (2012).

\bibitem{[BOR89]}
\newblock G.J.~Borjas,
\newblock {Economic theory and international migration},
\newblock {\em International Migration Review}, \textbf{23} 457-485, (1989).

\bibitem{[BW15]}
\newblock A.~Borzi and S.~Wongkaew,
\newblock Modeling and control through leadership of a refined flocking system
\newblock {\em Math. Models Methods Appl. Sci.}, \textbf{25} 565-585, (2015).

\bibitem{[BDG15]}
\newblock D.~Burini, S. De Lillo, and L.~Gibelli,
\newblock Collective learning dynamics modeling based on the kinetic theory of active particles,
\newblock {\em Phys. Life Rev.}, to appear.

\bibitem{[CAM03]}
\newblock C.F.~Camerer,
\newblock \textbf{Behavioral Game Theory: Experiments in Strategic Interaction},
\newblock Princeton University Press, (2003).

\bibitem{[CLP15]}
\newblock M.~Caponigro, A.C.~Lai, B.~Piccoli,
\newblock A nonlinear model of opinion formation on the sphere,
\newblock {\em Disc. Cont. Dyn. Syst. - Series A}, \textbf{35} 4241-4268, (2015).

\bibitem{[CFPT15]}
\newblock M.~Caponigro, M.~Fornasier, B.~Piccoli, E.~Tr\'elat
\newblock Sparse stabilization and control of alignment models
\newblock {\em Math. Models Methods Appl. Sci.}, \textbf{25} 561-564, (2015).

\bibitem{[CHEN13]}
\newblock J.~Chen \textit{et al.},
\newblock Synergystic challenges in data-intensive science and exascale computing,
\newblock  DOE ASCAC Data Sub-cmmittee Report, Office of Science, Department of Energy, USA, (2013).

\bibitem{[CPT05]}
\newblock S.~Cordier, L.~Pareschi, and G.~Toscani,
\newblock On a kinetic model for a simple market economy
\newblock {\em J. Stat. Phys.},  \textbf{120} 253-277, (2005).

\bibitem{[CPP09]}
\newblock S.~Cordier, L.~Pareschi, and C.~Piatecki,
\newblock Modelling financial markets,
\newblock {\em J. Stat. Phys.},  \textbf{134} 161-184, (2009).

\bibitem{[COU07]}
\newblock I.D.~Couzin,
\newblock {Collective minds},
\newblock {\em Nature},  \textbf{445} 715, (2007).

\bibitem{[DAC15]}
\newblock L.~D'Acci,
\newblock Mathematize {\em Urbes} by humanizing them. Cities as isobenefit landscapes: psycho-economical distances and personal isobenefit lines,
\newblock {\em Landscape and Urban Planning},  \textbf{139} 63-81, (2015)

\bibitem{[DL13]}
\newblock M.~Delitala and T.~Lorenzi,
\newblock A mathematical model for value estimate with public information and herding,
\newblock {\em Kinet. Related Mod.}, \textbf{7(1)} 29-44, (2014).

\bibitem{[DEA14]}
\newblock E.~De Angelis,
\newblock On the mathematical theory of post-Darwinian mutations, selection, and evolution,
\newblock {\em Math. Models Methods Appl. Sci.}, \textbf{24}  2723-2742, (2014).

\bibitem{[DPR04]} P.~Degond, L.~Pareschi, and G.~Russo,
\newblock \textbf{Modeling and Computational Methods for Kinetic Equations},
\newblock Series: Modeling and Simulation in Science, Engineering and Technology, Birkh\"auser, (2004).

\bibitem{[DSJ15]}
\newblock H.~De Sterck and C.~Johnson,
\newblock Data science: What is it and how is it thought?
\newblock \textit{SIAM News}, \textbf{48(6)} 1-6, (2015).

\bibitem{[DL14]}
\newblock M.~Dolfin and M.~Lachowicz,
\newblock Modeling altruism and selfishness in welfare dynamics: the role of nonlinear interactions,
\newblock  \emph{Math. Models Methods Appl. Sci.}, \textbf{24} 2361-2381, (2014).

\bibitem{[DMP12]}
B.~D\"uring, P.~Markowich, J.F. Pietschmann, and M.T.~Wolfram,
\newblock Boltzmann and {F}okker-{P}lanck equations modelling opinion formation  in the presence of strong leaders,
\newblock {\em P. R. Soc. London}, \textbf{465} 3687-3708, (2009).

\bibitem{[EMV06]}
\newblock G.C.~Ehrhardt,  M.~Marsili, and F.~Vega-Redondo,
\newblock {Phenomenological models of socioeconomic network dynamics},
\newblock {\em Physical Review E}, \textbf{74}, (2006),  paper n.~036106.

\bibitem{[FLL02]}
\newblock P.~Fajnzlber, D.~Lederman, and N.~Loayza,
\newblock {Inequality and violent crime},
\newblock  {\em J. Law Econ.}, \textbf{45} 1-40, (2002).

\bibitem{[FEL10]}
\newblock M.~Felson,
\newblock {What every mathematician should know about modelling crime},
\newblock  {\em Eur. J. Appl. Math.}, \textbf{21} 275-281, (2010).

\bibitem{[FFG09]}
\newblock A. Frezzotti, B. Franzelli, and L. Gibelli,
\newblock A moment method for low speed microflows,
\newblock {\em Continuum Mech. Thermodyn.}, \textbf{21} 495-509, (2009).

\bibitem{[FGG11a]}
\newblock A. Frezzotti, G. P. Ghiroldi, and L. Gibelli,
\newblock Solving model kinetic equations on GPUs,
\newblock {\em Comput. Fluids}, \textbf{50} 136-146, (2011).

\bibitem{[FGG11b]}
\newblock A. Frezzotti, G. P. Ghiroldi, and L. Gibelli,
\newblock Solving the Boltzmann equation on GPUs,
\newblock {\em Comput. Phys. Commun.}, \textbf{182} 2445-2453, (2011).

\bibitem{[GAL13]}
\newblock S.~Galam,
\newblock   \textbf{Sociophysics},
\newblock Springer, New York, (2013).

\bibitem{[GAL13A]}
\newblock S.~Galam and A.C.R.~Martins,
\newblock Building up of individual inflexibility in opinion dynamics
\newblock {\em Physical Review E}, \textbf{87} paper n.~042807, (2013).

\bibitem{[GL05]}
\newblock D.~Garlaschelli and M.I.~Loffredo,
\newblock Structure and evolution of the world trade network,
\newblock  {\em Physica A}, \textbf{355} 138-144, (2005).

\bibitem{[GL08A]}
\newblock D.~Garlaschelli and M.I.~Loffredo,
\newblock Maximum likelihood: Extracting unbiased information from complex networks,
\newblock  {\em Phys. Rev. E}, \textbf{78} paper n.~015101, (2008).

\bibitem{[GL08B]}
\newblock D.~Garlaschelli and M.I.~Loffredo,
\newblock Effect of network topology on wealth distribution,
\newblock  {\em J. Phys. A: Math. Theor.}, \textbf{41} paper n.~2240018, (2008).

\bibitem{[GAL13B]}
\newblock S.~Galam,
\newblock The drastic outcomes from voting alliances in three-party democratic voting,
\newblock  {\em J. Statist. Phys.}, \textbf{151} 46-68, (2013).

\bibitem{[GG14]}
\newblock G.P.~Ghiroldi  and L.~Gibelli,
\newblock A direct method for the Boltzmann equation based on a pseudo-spectral velocity space discretization,
\newblock \emph{J. Comput. Phys.}, \textbf{258} 568-584, (2014).

\bibitem{[G12]}
\newblock L.~Gibelli,
\newblock Velocity slip coefficients based on the hard-sphere Boltzmann equation,
\newblock \emph{Phys. Fluids}, \textbf{24} paper n.~022001 (2012).

\bibitem{[GP09]}
\newblock F.~Gino and L.~Pierce,
\newblock The abundance effect: Unethical behavior in the presence of wealth,
\newblock {\em Organ. Behav. Hum.}, \textbf{109} 142-155, (2009).

\bibitem{[GIN09]}
\newblock H.~Gintis,
\newblock \textbf{Game Theory Evolving}, 2nd Ed.,
\newblock Princeton University Press, Princeton NJ, (2009).

\bibitem{[GRO12]}
\newblock M.~Gromov, In a search for a structure: on entropy,\\
\newblock   www.ihes.fr/~gromov/PDF/structre-serch-entropy-july5-2-2012.pdf, July (2012).

\bibitem{[HHM10]}
\newblock S.~Harrendorf, M.~Heiskanen and S.~Malby,
\newblock \emph{International statistics on crime and justice},
\newblock European Institute for Crime Prevention and Control, affiliated with the United Nations (HEUNI), (2010).

\bibitem{[HK02]}
\newblock R.~Hegselmann and U.~Krause,
\newblock {Opinion dynamics and bounbded confidence: Models analysis and social simulations},
\newblock {\em JASS-The J. of Artificial Society and Social Simulations}, \textbf{5(3)}, (2002).

\bibitem{[HK15]}
\newblock R.~Hegselmann and U.~Krause,
\newblock {Opinion dynamics under the influence of radical groups, charismatic and leaders, and other constant signals: A simple unifying model}
\newblock {\em Netw. Het. Media}, (2015), to be published.

\bibitem{[HEL10]}
\newblock D.~Helbing,
\newblock  \textbf{Quantitative Sociodynamics. Stochastic Methods and Models of  Social Interaction Processes},
\newblock Springer Berlin Heidelberg, 2nd edition, (2010).

\bibitem{[HY09]}
\newblock D.~Helbing and W.~Yu
\newblock The outbreak of cooperation among success-driven individuals under noisy conditions,
\newblock {\em Proc. Natl. Acad. Sci. USA},  \textbf{106} 3680-3685, (2009).

\bibitem{[HER11]}
\newblock M.A.~Herrero,
\newblock Through a glass, darkly: biology seen from mathematics: comment on ``Toward a mathematical theory of living systems focusing on developmental biology and evolution: a review and perspectives'' by Bellomo N. and Carbonaro B.,
\newblock {\em Physics of Life Reviews},  \textbf{8} 21, (2011).

\bibitem{[HS03]}
\newblock J.~Hofbauer and K.~Sigmund,
\newblock Evolutionary game dynamics,
\newblock {\em Bull. Am. Math. Society},  \textbf{40} 479-519, (2003).

\bibitem{[HP93]}
\newblock C.C.~Hsieh and M.D.~Pugh,
\newblock  Poverty, Income inequality, and violent crime: a meta-analysis of recent aggregate data studies,
\newblock {\em Crim. Just. Rev.}, \textbf{18} 182-202, (1993).

\bibitem{[REPORT]}
\newblock http://www.cs.utah.edu/bigdata/,
\newblock http://www.science.vt.edu/ais/cmda/,
\newblock http://www.analytics/gatech.edu/.

\bibitem{[KA12]}
\newblock D.~Kahneman,
\newblock \textbf{Thinking Fast and Slow},
\newblock Penguin Books, London, (2012).

\bibitem{[KIR11]}
\newblock A.~Kirman
\newblock \textbf{Complex Economics: Individual and Collective Rationality},
\newblock Routledge, London, (2011).

\bibitem{[KZ01]}
\newblock A.P.~Kirman and J.B.~Zimmermann,
\newblock \textbf{Economics with Heterogeneous Interacting Agents},
\newblock {Lecture Notes in Economics and Mathematical Systems}, \textbf{503}, Springer, (2001).

\bibitem{[KNO13]}
\newblock D.~Knopoff,
\newblock On the modeling of migration phenomena on small networks,
\newblock \emph{Math. Models Methods Appl. Sci.}, \textbf{23} 541-563, (2013).

\bibitem{[KNO14]}
\newblock D.~Knopoff,
\newblock On a mathematical theory of complex systems on networks with application to opinion formation,
\newblock \emph{Math. Models Methods Appl. Sci.}, \textbf{24} 405-426, (2014).

\bibitem{[KSAX10]}
\newblock M. Kolar, L. Song, A. Ahmed and E.P. Xing,
\newblock Estimating time-varying networks,
\newblock {\em The Annals of Applied Statistics}, \textbf{4} 94-123, (2010).

\bibitem{[KPK09]}
\newblock M.W.~Kraus, P.K.~Piff, and D.~Keltner,
\newblock Social class, sense of control, and social explanation,
\newblock {\em J. Pers. Soc.  Psychol.}, \textbf{99} 771-784, (2009).

\bibitem{[KPM12]}
\newblock M.W.~Kraus, P.K.~Piff, R.~Mendoza-Denton, M.L.~Rheinschmidt, and D.~Keltner,
\newblock  Social class, solipsism, and contextualism: How the rich are different from the poor,
\newblock {\em Psychological Review}, \textbf{119} 546-572, (2012).

\bibitem{[LL07]}
\newblock J.M.~Lasry and P.L.~Lions,
\newblock {Mean field games},
\newblock {\em Japan. J. Math.}, \textbf{2} 229-260, (2007).

\bibitem{[MAY04]}
\newblock R.M.~May,
\newblock Uses and abuses of mathematics in biology,
\newblock  {\em Science},  \textbf{303} 790-793, (2004).

\bibitem{[MAYR70]}
\newblock E.~Mayr,
\newblock   \textbf{Populations, Species, and Evolution},
\newblock Harward University Press, (1970).

\bibitem{[MAYR01]}
\newblock E.~Mayr,
\newblock   \textbf{What Evolution Is},
\newblock Basic Books, New York, (2001).

\bibitem{[MT14]}
\newblock S.~Motsch and E.~Tadmor
\newblock  Heterophilious dynamics enhances consensus,
\newblock {\em SIAM Review}, \textbf{56} 577-621, (2014).

\bibitem{[NASH1]}
\newblock J.~Nash,
\newblock Noncoperative games,
\newblock  {\em Ann. of Mathematics}, \textbf{54} 287-295, (1951).

\bibitem{[NASH2]}
\newblock J.~Nash,
\newblock \textbf{Essentials of Game Theory},
\newblock Elgar, Cheltenham, (1996).

\bibitem{[NOW06]}
\newblock M.A.~Nowak,
\newblock  \textbf{Evolutionary Dynamics. Exploring the Equations of Life}.
\newblock Harvard University Press, (2006).

\bibitem{[NHP11]}
J.~C. Nu\~{n}o, M.~A. Herrero, and M.~Primicerio.
\newblock A mathematical model of criminal-prone society.
\newblock {\em Discrete Contin. Dyn. Syst. Ser. S}, \textbf{4(1)} 193-207, (2011).

\bibitem{[ORM05]}
\newblock P.~Ormerod,
\newblock \emph{Crime: Economic incentives and social networks},
\newblock IEA Hobart Paper 151, (2005).

\bibitem{[PT13]}
\newblock L.~Pareschi and G.~Toscani,
\newblock  \textbf{Interacting Multiagent Systems: Kinetic Equations and Monte Carlo Methods}
\newblock Oxford University Press, Oxford, (2013).

\bibitem{[PSC14]}
\newblock P.K.~Piff, D.M.~Stancato, S.~Cot\'e, R.~Mendoza-Denton, and D.~Keltner,
\newblock Higher social class predicts increased unethical behavior,
\newblock {\em Proc. Natl. Acad. Sci. USA},  \textbf{109} 4086-4091, (2014).

\bibitem{[RAC11]}
\newblock D.~G. Rand, S.~Arbesman, and N.~A. Christakis,
\newblock Dynamic social networks promote cooperation in experiments with humans,
\newblock {\em Proc. Natl. Acad. Sci. USA},  \textbf{108(48)} 19193-19198, (2011).

\bibitem{[RBE12]}
\newblock P.~Romanczuk, M.~Bar, W.~Ebeling, B.~Lindner, and L.~Schimansky-Geier,
\newblock Active Brownian particles, from individual to collective stochastic dynamics,
\newblock {\em The Eurpean Physical Journal}, \textbf{202} 1-162, (2012).

\bibitem{[SPL06]}
\newblock F.C.~Santos, J.M.~Pacheco, and T.~Lenaerts,
\newblock Evolutionary dynamics of social dilemmas in structured heterogeneous populations,
\newblock {\em Proc. Natl. Acad. Sci. USA},  \textbf{103(9)} 3490-3494, (2006).

\bibitem{[SVS12]}
\newblock F.C.~Santos, V.~Vasconcelos, M.D.~Santos, P.~Neves, and J.M.~Pacheco,
\newblock Evolutionary dynamics  of climate change under collective-risk dilemmas,
\newblock {\em Math. Models Methods Appl. Sci.},  \textbf{22}, (2012), paper n.~1140004.

\bibitem{[SBB09]}
\newblock M.~Scheffer, J.~Bascompte, W.A.~Brock, V.~Brovkin, S.R.~Carpenter, V.~Dakos, H.~Held, E.H.~van Nes, M.~Rietkerk, and G.~Sugihara,
\newblock Early-warning signals for critical transitions.
\newblock {\em Nature},  \textbf{461} 53-59, (2009).

\bibitem{[SBD10]}
\newblock M.~B.~Short, P.~J.~Brantingham, and M.~R.~D'Orsogna,
\newblock {Cooperation and punishment in an adversarial game: how defectors pave the way to a peaceful society},
\newblock {\em Phys. Rev. E}, \textbf{82} (2010), paper n.~066114.

\bibitem{[SDP08]}
\newblock M.B.~Short, M.R.~D'Orsogna, V.B.~Pasour, G.E.~Tita, P.J.~Brantingham, A.L.~Bertozzi, and L.B.~Chayes,
\newblock {A statistical model of criminal behavior},
\newblock {\em Math. Models Methods Appl. Sci.}, \textbf{18} 1249-1267, (2008).

\bibitem{[SIG11]}
\newblock K.~Sigmund
\newblock \textbf{The {C}alculus of {S}elfishness}
\newblock {Princeton University Series in Theoretical and Computational Biology}, Princeton, USA, (2011).

\bibitem{[SIM59]}
\newblock H.A.~Simon,
\newblock Theories of decision-making in {E}conomics and {B}ehavioral {S}cience,
\newblock {\em Am. Econ. Rev.},  \textbf{49(3)} 253-283, (1959).

\bibitem{[SIM97]}
\newblock H.A.~Simon,
\newblock   \textbf{Models of Bounded Rationality},
\newblock MIT Press, Vols.1-2., (1982), Vol.3, (1997).

\bibitem{[STIG09]}
\newblock J.E.~Stiglitz,
\newblock {Information and the change in the paradigm in economics},
\newblock {\em The American Economic Review},  \textbf{92} 460-501, (2009).

\bibitem{[ST14]}
\newblock B.~Straughan,
\newblock Shocks and acceleration waves in modern continuum machanics and in social systems,
\newblock {\em Evol. Eq. Control Theory}, \textbf{3}, 541-555, (2014).

\bibitem{[TAL07]}
\newblock N.N.~Taleb,
\newblock   \textbf{The Black Swan: The Impact of the Highly Improbable},
\newblock Random House, New York City, (2007).

\bibitem{[TKA74]}
\newblock A.~Tversky and D.~Kahneman,
\newblock Judgment under uncertainty: Heuristics and biases,
\newblock {\em Science, New Series}, \textbf{185} 1124-1131, (1974).

\bibitem{[TOS06]}
\newblock G.~Toscani,
\newblock {Kinetic models of opinion formation},
\newblock {\em Comm. Math. Sci.}, \textbf{4} 481-496, (2006).

 \bibitem{[VR07]}
\newblock V.~Vega-Redondo,
\newblock \textbf{Complex Social Networks}.
\newblock Cambridge University Press, (2007).

\bibitem{[W92]}
\newblock W.~Wagner,
\newblock {A convergence proof of Bird's direct simulation Monte-Carlo method for the Boltzmann equation},
\newblock {\em J. Stat. Phys.},  \textbf{66} 1011-1044, (1992).

\end{thebibliography}
\end{document}